\documentclass[12pt]{article} 
\usepackage{graphicx}
\usepackage{amsmath,amssymb}
\usepackage{longtable}

\begin{document}
\baselineskip 21pt

\bigskip

\centerline{\Large \bf Noncircular Outer Disks in Unbarred S0 Galaxies:}
\centerline{\Large \bf NGC 502 and NGC 5485}
\bigskip

\centerline{\large O. K. Sil'chenko}

\noindent
{\it Sternberg Astronomical Institute of the Lomonosov Moscow State University, Moscow, Russia}

\vspace{2mm}
\sloppypar 
\vspace{2mm}

\bigskip

{\small \bf
\noindent
Strongly noncircular outer stellar disks have been found in two unbarred SA0 galaxies by means of
analyzing spectroscopic data on the rotation of stars and photometric data on the shape and
orientation of the isophotes. In NGC~502, the oval distortion of the disk is manifested as two
elliptical rings, the inner and the outer ones, covering wide radial zones between the bulge and the disk and at the
outer edge of the stellar disk. Such a structure may be a consequence of the so-called ``dry'' minor merger -- multiple
accretion of gas-free satellites. In NGC~5485, the kinematical major axis does not coincide with the
orientation of isophotes in the disk-dominated region, and for this galaxy the conclusion about its global triaxial
structure is unavoidable.
}

\clearpage

\section{INTRODUCTION}

The main part of galaxies in the nearby Universe are disk galaxies. It means that one
of the large-scale components of their structure, often dominant, is just a stellar disk. From a dynamical
point of view, stellar disks are ``cold'' systems, i.e., chaotic motions of stars there are small, and
the bulk of their kinetic energy is confined to laminar circular rotation around the galactic center.
Round lines of equal density (or of equal surface brightness) correspond to circular orbits, and, on the whole, 
the galactic disk, when viewed pole-on, looks like a regular circle. But how perfect is a circular shape
of real galactic disks? This question worried researchers since the very rise of the
extragalactic astronomy. The problem is complicated by the fact that we see any disk of a
galaxy in projection onto the plane of the sky, at an arbitrary inclination, and it is not easy to 
distinguish the visible ellipticity produced by projection effects from the actual ``intrinsic'' disk 
ellipticity, especially in the case where we have to deal with individual objects and not with large samples. 
The average intrinsic ellipticity of stellar disks can be estimated statistically by studying their distribution 
over apparent isophote ellipticity under the assumption of a random orientation in space. In addition, since 
the assumption of circular gas rotation within the disks of galaxies is widely accepted while establishing the so-called 
scaling relations, for example, the Tully–Fisher relation, the scatter of galaxies observed around such relations 
also allows the intrinsic ellipticity of the disks to be constrained. Franx and de Zeeuw \cite{franx_dz92} had collected 
the photometric statistics and the statistics on the Tully–Fisher relation and had placed an upper limit onto 
the typical ellipticity of galactic disks, $\epsilon < 0.1$, by pointing out that it most likely lies within 
the range 0--0.06.

The paper by Rix and Zaritsky \cite{rix_zar95} can be mentioned as the first special work dedicated to this
topic and based on highly accurate surface photometry of the disks of several individual galaxies. They
had selected 18 galaxies seen nearly face-on judging that the rotation velocity projected onto the line of sight
in the flat disks of such galaxies must be zero and, consequently, the integrated 21-cm HI emission line
must be very narrow. The $K$--band surface photometry was then performed to minimize the effect of dust and
of the clumpy distribution of star-forming regions. The subsequent azimuthal Fourier analysis of the surface
brightness distribution proved the disks to be actually rather circular, with a mean intrinsic ellipticity
of 4.5\%. The fact that such studies were impossible without invoking kinematic data when individual
galaxies rather than average statistics were involved was obvious from the beginning, and new opportunities
to solve the problem of the intrinsic ellipticity of galactic disks appeared with the development of
integral-field spectroscopy. Andersen et al. \cite{andersen} calculated the intrinsic disk ellipticity
for seven Sb--Sc galaxies oriented nearly face-on by using both highly accurate red-band surface photometry
and two-dimensional line-of-sight velocity fields obtained with the DensePak integral-field fiber
spectrograph of the WIYN telescope. The true inclinations of the flat disks to the line of sight were determined 
from the two-dimensional ionized-gas kinematics. Their paper has a reference to the theoretical conclusion  
by Franx and de Zeeuw \cite{franx_dz92} that the method of tilted circular {\it(sic!)} rings can be applied 
to gas on elliptical orbits if the orbit ellipticity is small and if the rotation curve (the dependence of 
the rotation velocity on galactocentric distance) is flat. The conclusion about intrinsic ellipticity of the 
stellar disks was then derived from the shape of the outer isophotes. Its mean value turned out to be about 
5\%, in full agreement with the expectations, though the intrinsic disk ellipticity in two particular galaxies 
with close satellites which are possibly impacted by gravitational tides reached 20\%.

In this paper, we consider two lenticular galaxies for which our data on the kinematics of the stellar components 
have turned out to be incompatible with the assumption about circular shape and circular rotation of the disks. 
In the second section, we describe the galaxies under consideration and the observational data used here. The evidence
for the ellipticity of their stellar disks is formulated in the third section.

\section{NGC~502 AND NGC~5485: NONINTERACTING S0 GALAXIES IN GROUPS}

Table~1 provides the basic parameters of the galaxies under consideration which are accumulated in catalogs.
Both targets are intermediate-luminosity lenticular galaxies; they are members of galaxy groups. However,
while NGC~5485 is probably the central galaxy in its group \cite{huchra_geller,garcia},
NGC~502 is located on the periphery of its group, approximately at 300 kpc from the group central galaxy NGC~524,
and only very faint satellites can be seen around it, within the radius of 100--150~kpc \cite{n524phot}.

The galaxies to be discussed entered into a large sample of lenticular galaxies that we investigated by means of
integral-field spectroscopy at the 6-m telescope of the Special Astrophysical Observatory belonging to the Russian
Academy of Sciences (SAO RAS) \cite{sil08}. In particular, two-dimensional velocity maps of the
stellar components were constructed for the central regions. A misalignment of the kinematical and photometric
major axes was detected even when the stellar kinematics of the innermost regions of NGC~502 and NGC~5485 
was considered. In principle, this is not quite uncommon for the central regions of galaxies, 
but it is usually accompanied (and explained) by the presence of a bar at the galactic center violating 
axial symmetry in the gravity potential distribution and, accordingly, producing the consequent departure
from a circular character of rotation. However, NGC~502 and NGC~5485 do not have not only bars but 
even hints of an oval bulge distortion; both galaxies are classified as SA0. We became interested in the
unusual kinematics of the stellar components in these galaxies and undertook long-slit spectroscopy 
to see what happens in the outermost, disk-dominated regions of the galaxies. The slit was aligned 
with the major axes of the outer isophotes that must have coincided with the line of nodes of
the disk plane under the assumption of the circular disk shape. To our great surprise, our spectroscopy with
a long slit oriented in the presumed direction of the maximum rotation velocity projected onto the line of
sight showed no rotation in NGC~5485 and a very weak one in NGC~502! We decided to collect all necessary 
information to figure out what causes this phenomenon.

\begin{table}
\caption{Global parameters of the galaxies}
\begin{tabular}{l||p{3cm}|p{3cm}|}
\hline
NGC & 502 & 5485 \\
\hline
Morphological type (NED$^1$) & SA(r)0$^0$ & SA0 pec \\
Distance, Mpc (NED) & 34 & 27 \\
$R_{25}$ (RC3$^2$) & $34''$ & $70''$ \\
$R_{25}$, kpc & 6.0 & 9.5 \\
$B_T^0$ (RC3) & 13.57 & 12.31 \\
$M_B$ (LEDA$^3$) & --19.3 & --20.2 \\
$(B-V)^0_T$ (RC3) & 0.91 & 0.88 \\
$(U-B)^0_T$ (RC3) & 0.47 & 0.51 \\
$V_r$ (NED), km s$^{-1}$ & 2524 & 1927 \\
Inclination (LEDA) & $24^{\circ}$ & $55^{\circ}$ \\
$PA_{phot}$ ($R_{25}$) & $60^{\circ}$$^4$ & $170^{\circ}$$^2$ \\
$\sigma_*$, km s$^{-1}$ (LEDA) & 103 & 195 \\
\hline
\multicolumn{3}{l}{$^1$\rule{0pt}{11pt}\footnotesize
NASA/IPAC Extragalactic Database.}\\
\multicolumn{3}{l}{$^2$\rule{0pt}{11pt}\footnotesize
Third Reference Catalogue of Bright Galaxies \cite{rc3}. }\\
\multicolumn{3}{l}{$^3$\rule{0pt}{11pt}\footnotesize
Lyon–Meudon Extragalactic Database.}\\
\multicolumn{3}{l}{$^4$\rule{0pt}{11pt}\footnotesize
Il'ina and Sil'chenko \cite{n524phot}.}\\
\end{tabular}
\end{table}

\begin{table}
\caption{Long-slit spectroscopy of the galaxies}
\begin{tabular}{rlccc}
\hline
Galaxy & Date & T(exp), min & $PA_{slit}$ & FWHM, arcsec \\
NGC 502 & Sep. 3, 2008 & 80 & 64 & 2.5 \\
NGC 502 & Sep. 3, 2008 & 80 & 334 & 1.6 \\
NGC 502 & Nov. 3, 2010 & 60 & 30 & 1.2 \\
NGC 502 & Nov. 3, 2010 & 120 & 15 & 1.2 \\
NGC 5485 & May 12, 2010 & 60 & 14 & 2.2 \\
NGC 5485 & Feb. 9, 2011 & 120 & 75 & 2.3 \\
NGC 5485 & Mar. 19, 2015 & 90 & 120 & 2.0 \\
\hline
\end{tabular}
\end{table}

\begin{figure*}[p]
\resizebox{\hsize}{!}{\includegraphics{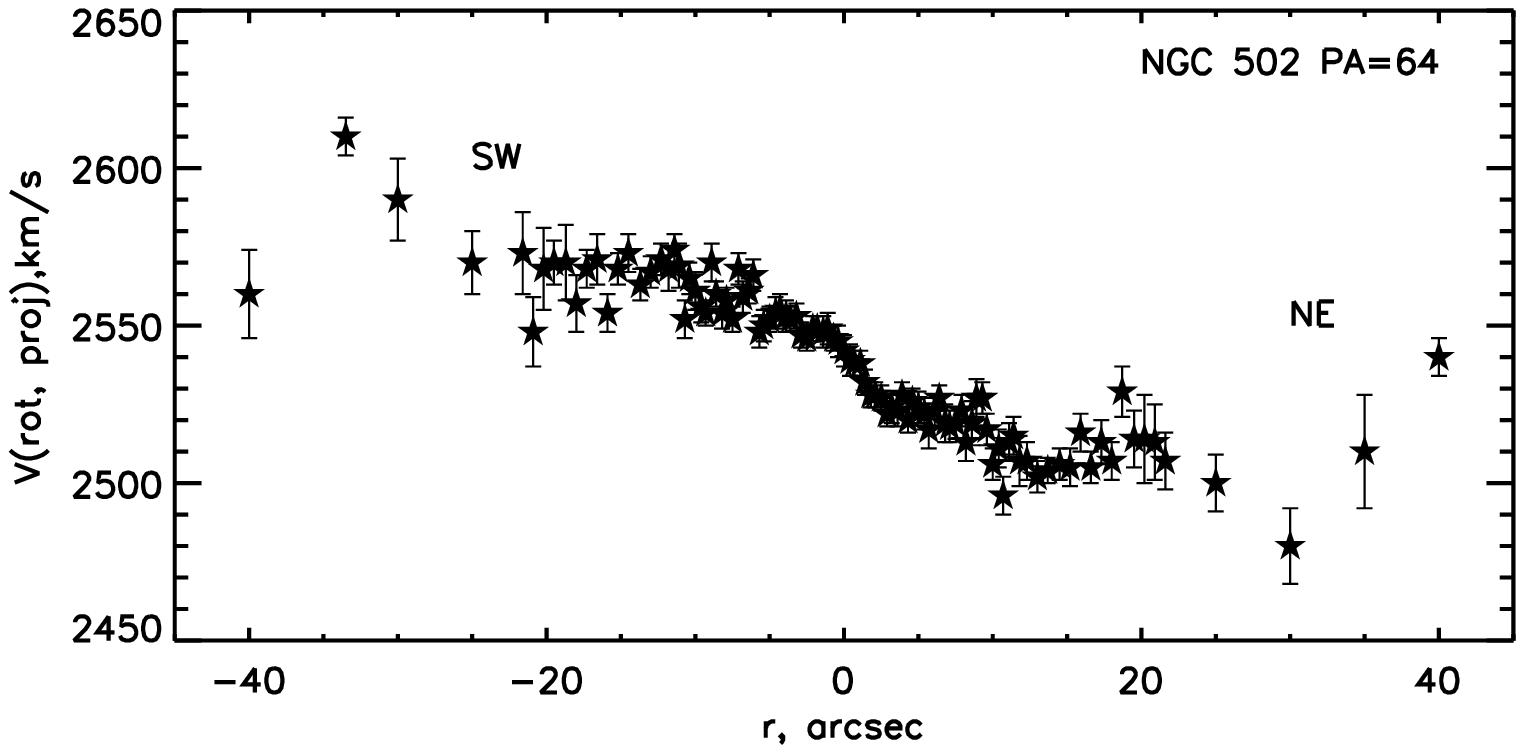}
\includegraphics{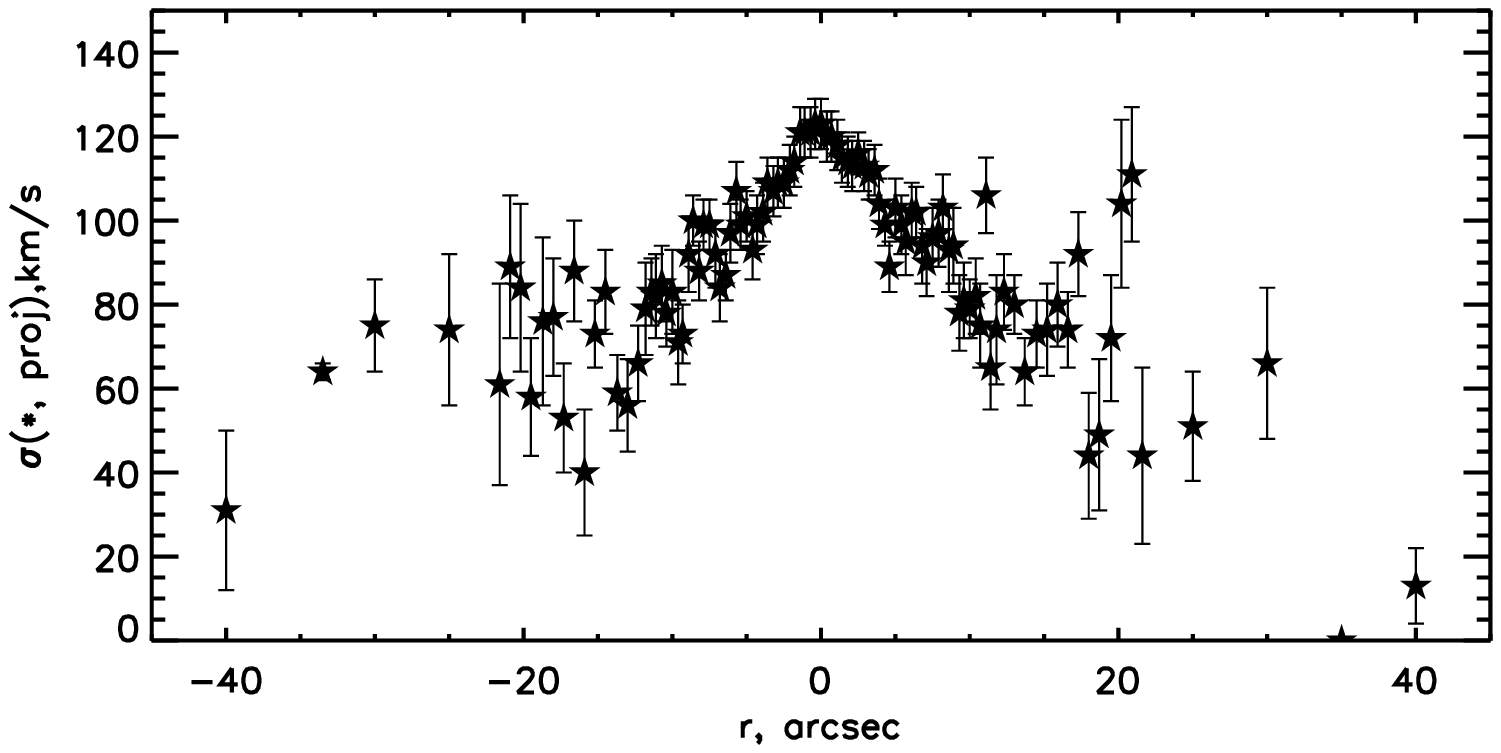}}
\resizebox{\hsize}{!}{\includegraphics{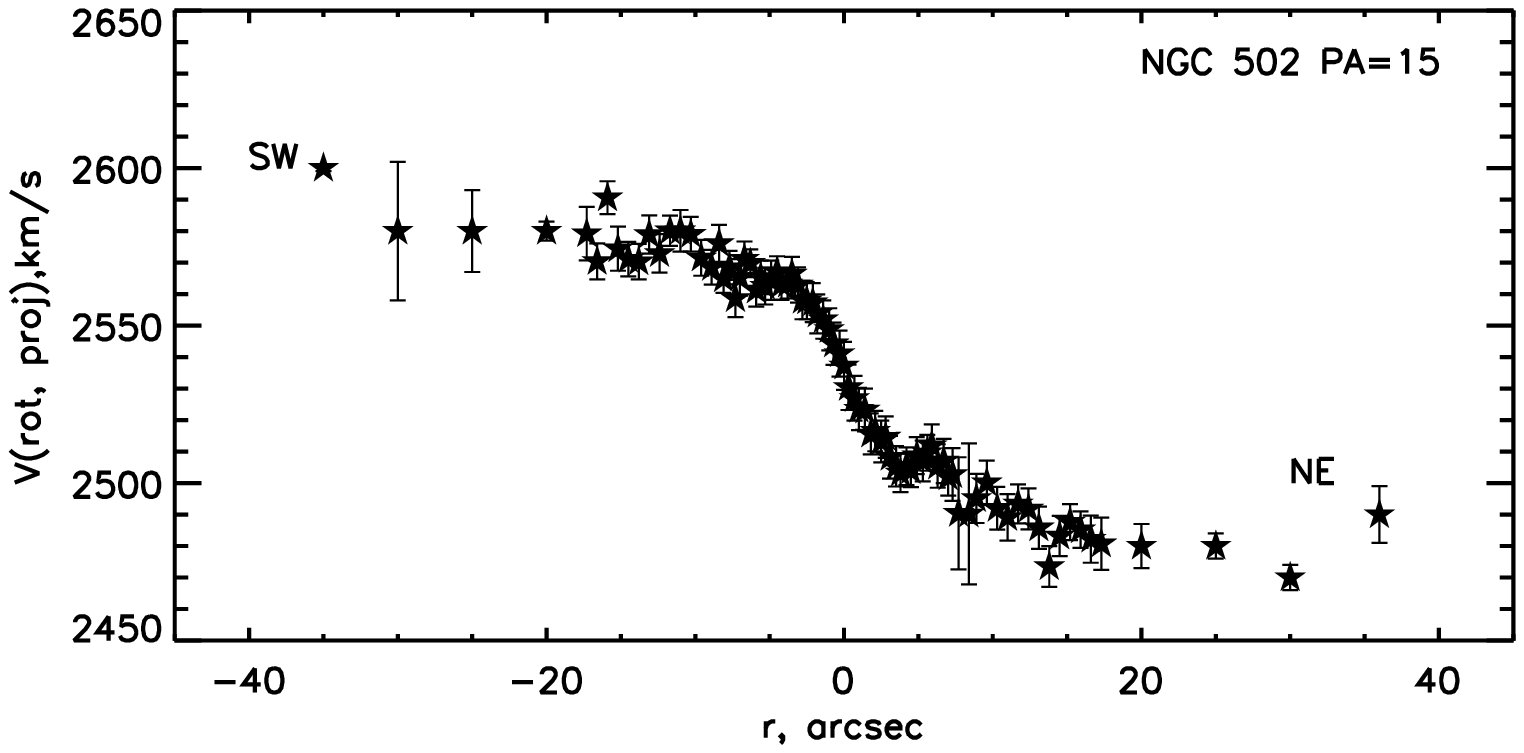}
\includegraphics{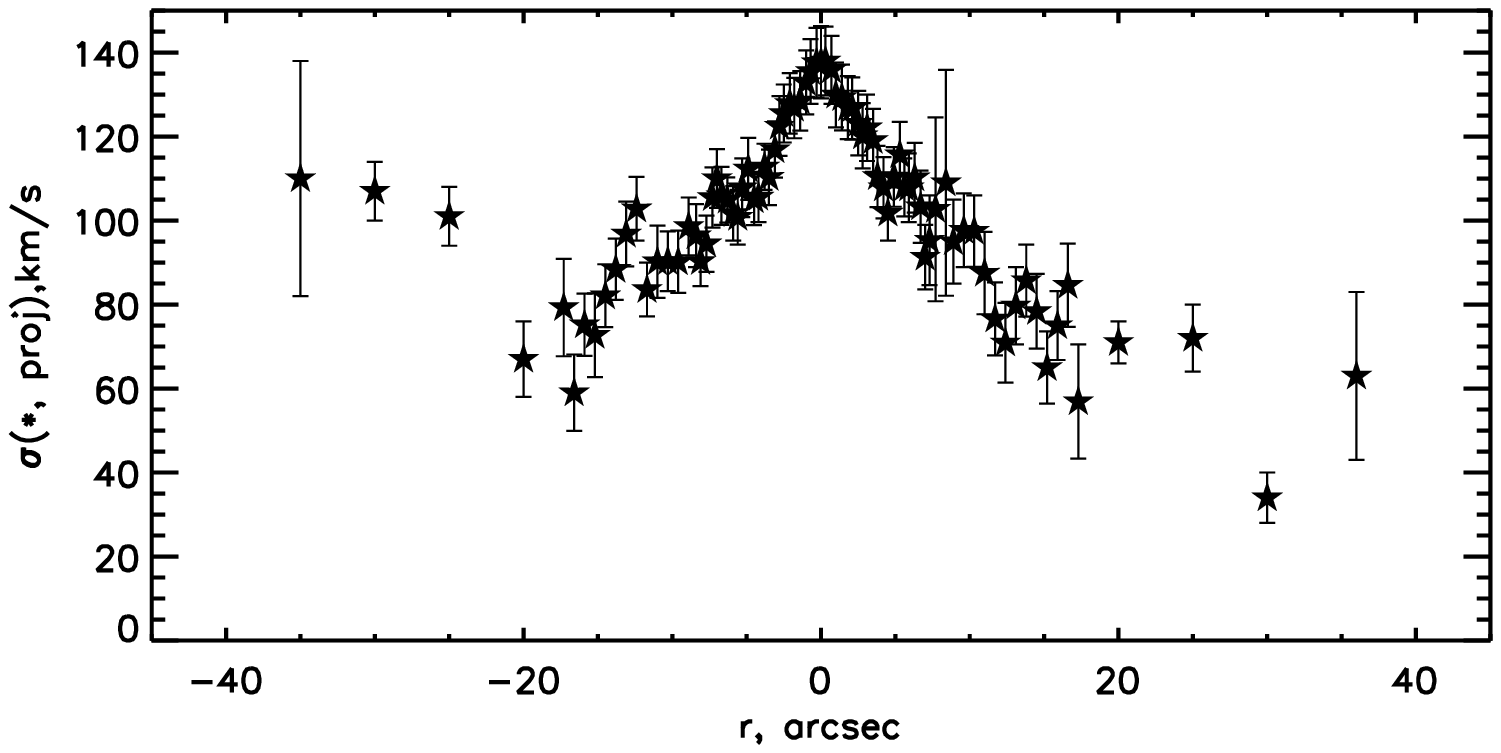}}
\resizebox{\hsize}{!}{\includegraphics{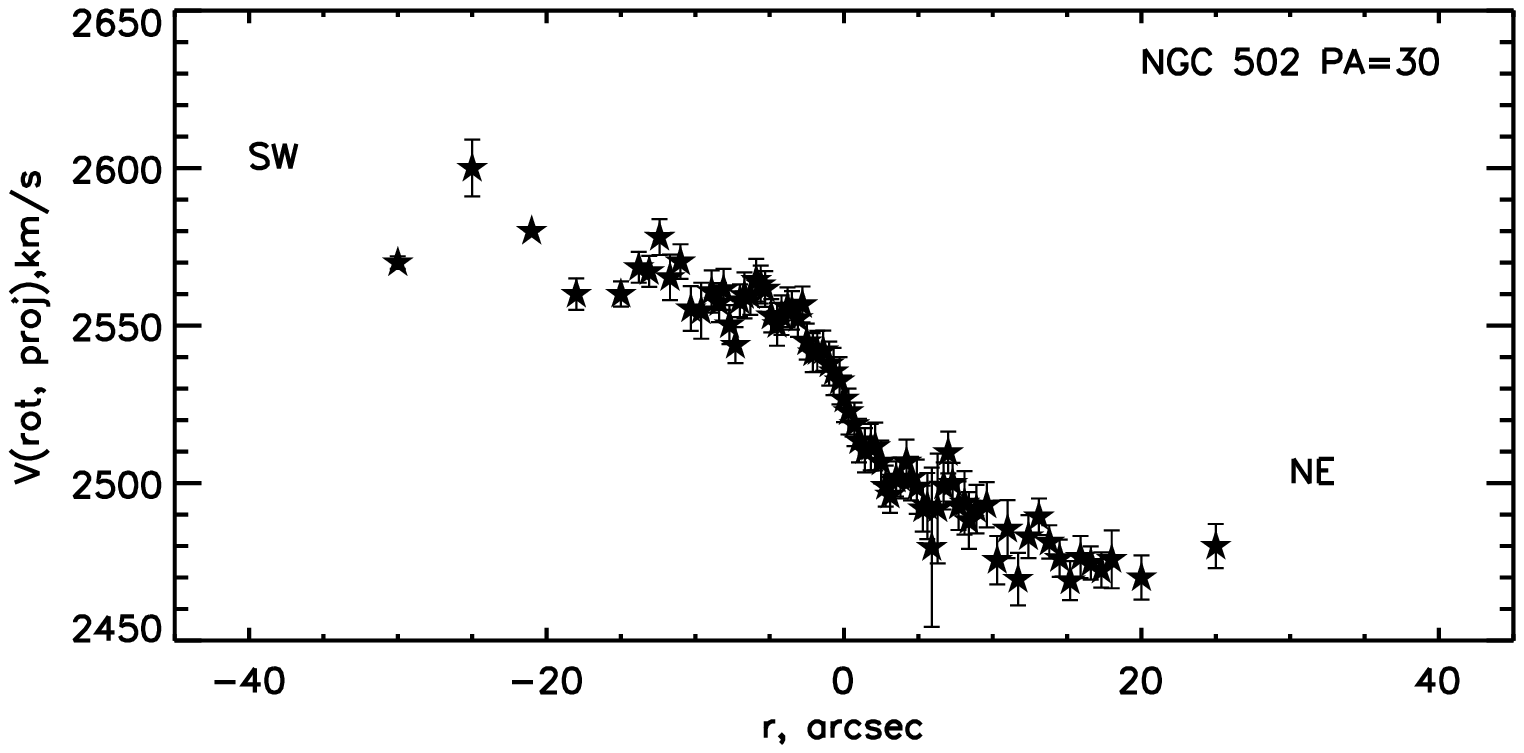}
\includegraphics{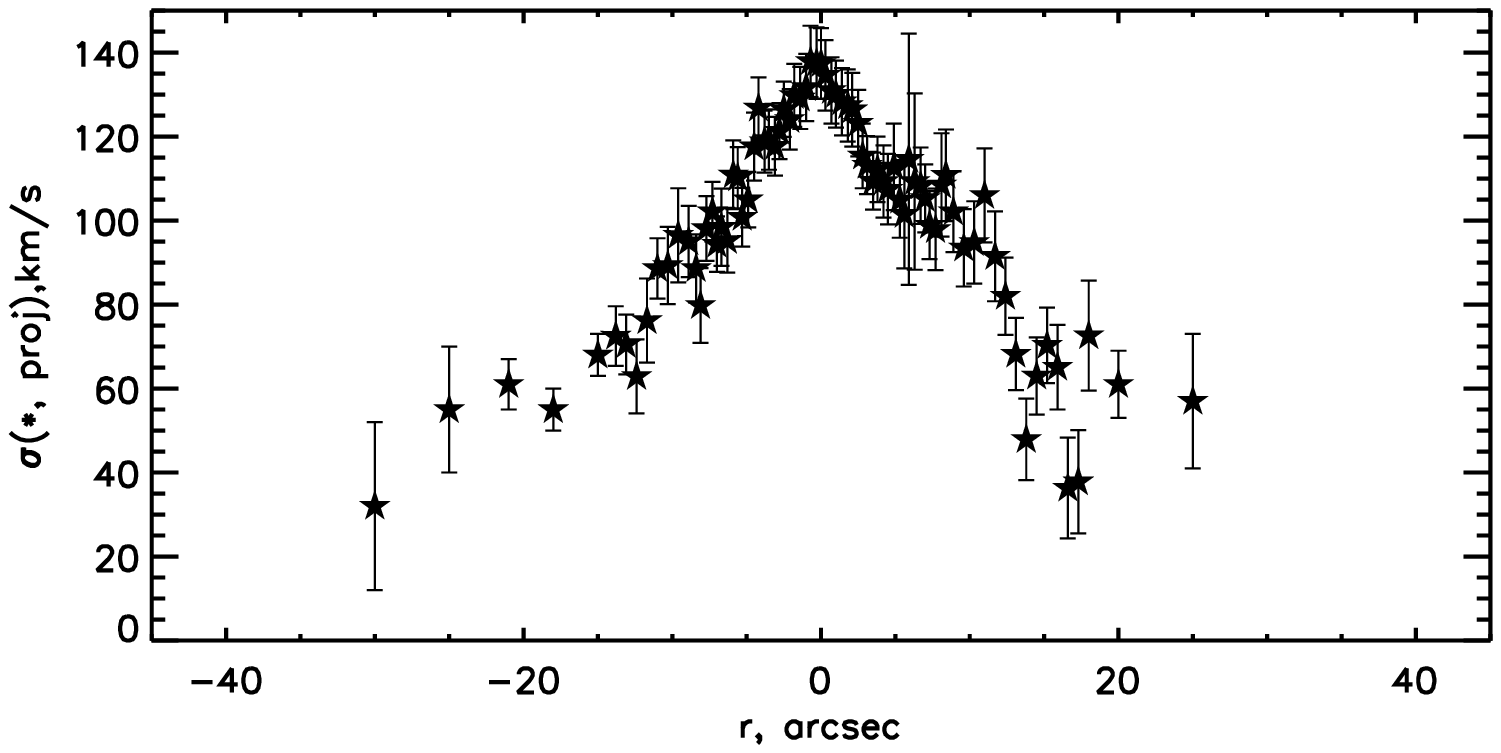}}
\resizebox{\hsize}{!}{\includegraphics{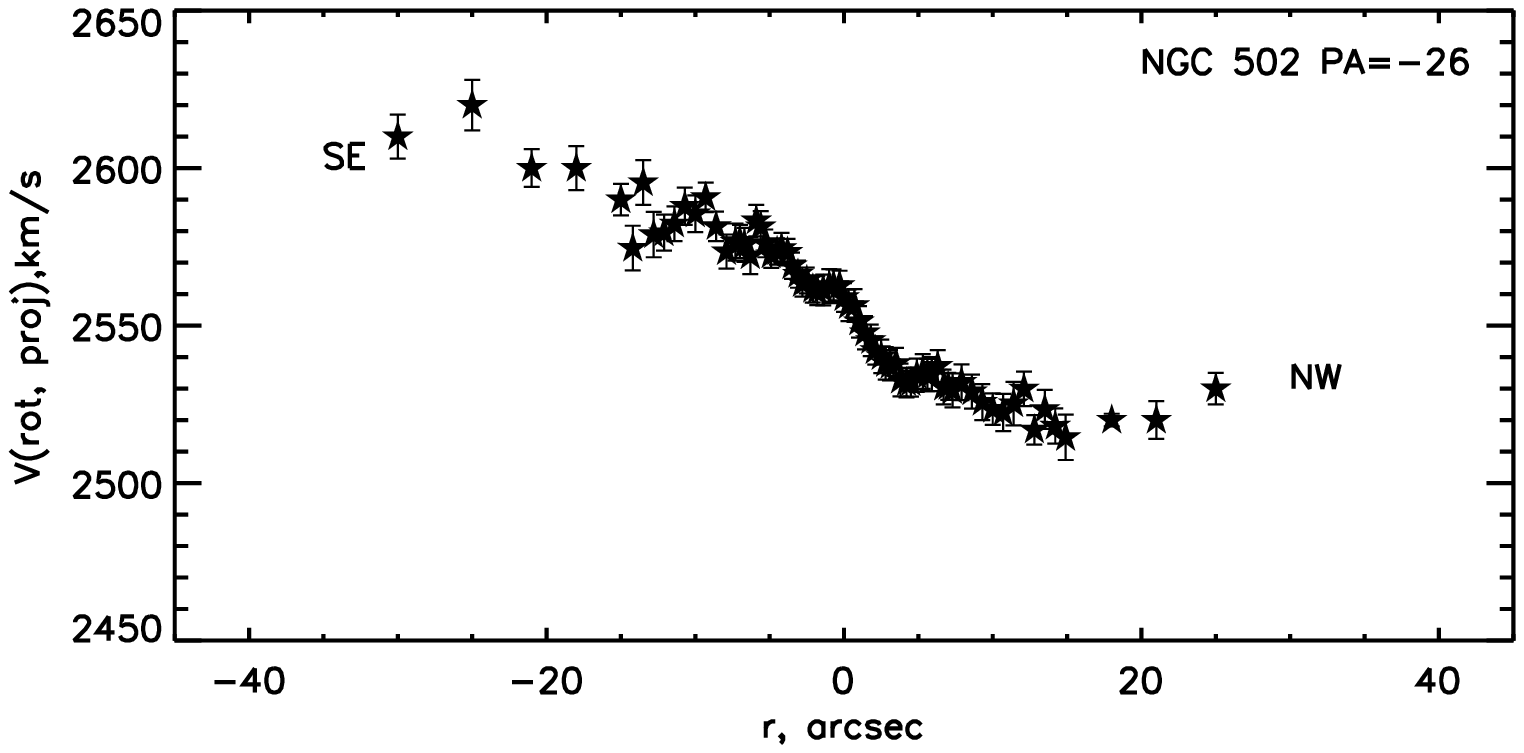}
\includegraphics{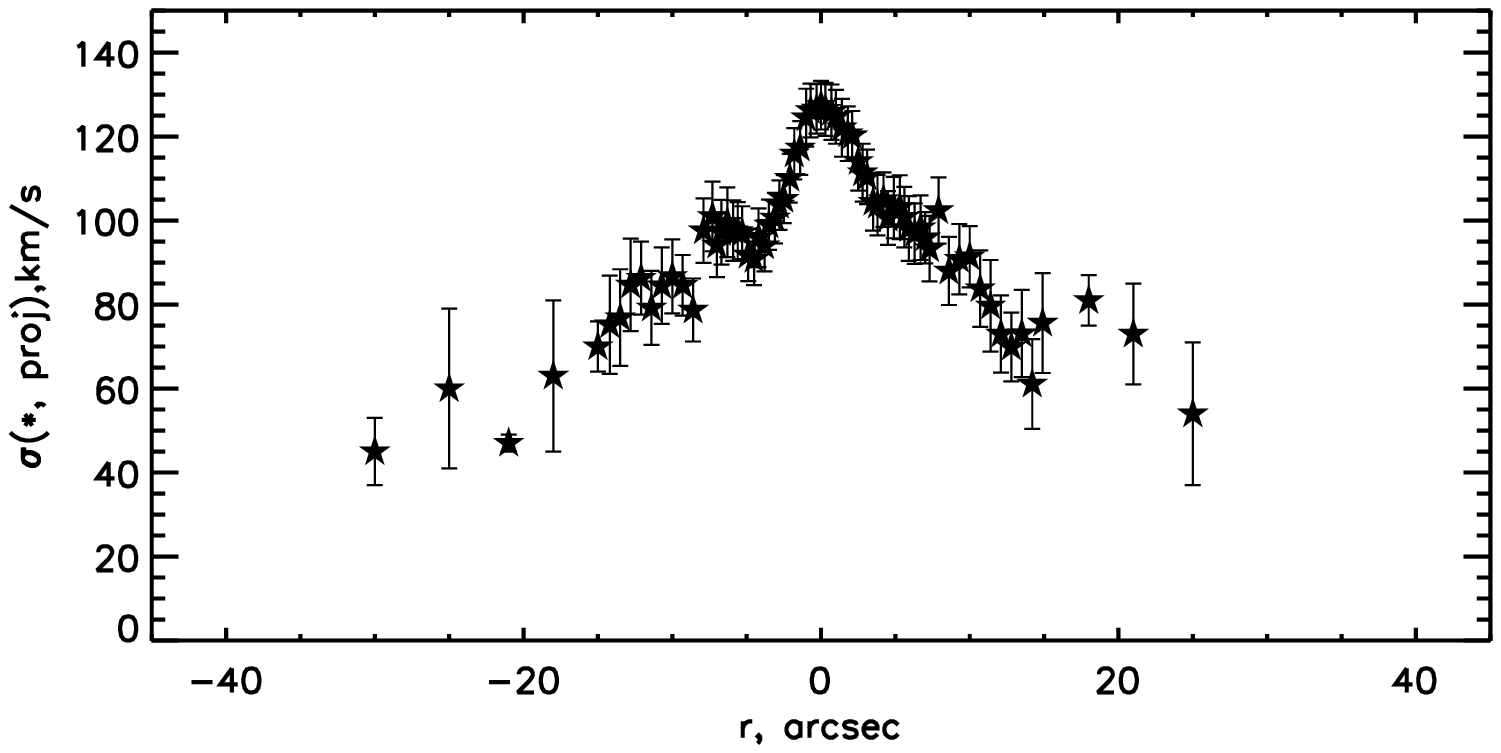}}
\caption{Kinematical cross-sections obtained for NGC~502 with the SCORPIO spectrograph in the long-slit mode 
at the 6-m SAO RAS telescope at four different position angles: the stellar line-of-sight velocities ({\it left}) 
and the corresponding stellar velocity dispersions ({\it right}).}
\label{n502longslit}
\end{figure*}

\begin{figure*}[p]
\resizebox{\hsize}{!}{\includegraphics{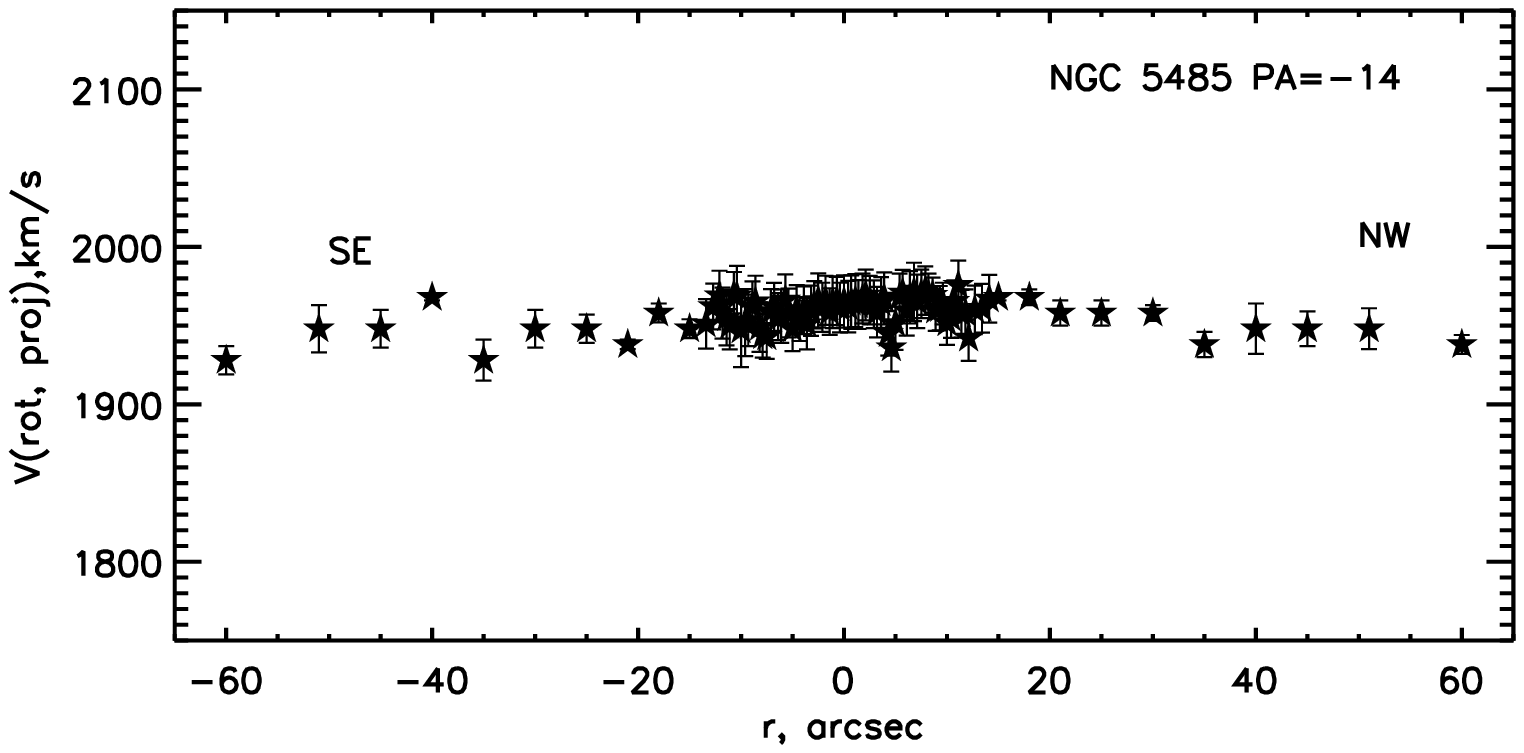}
\includegraphics{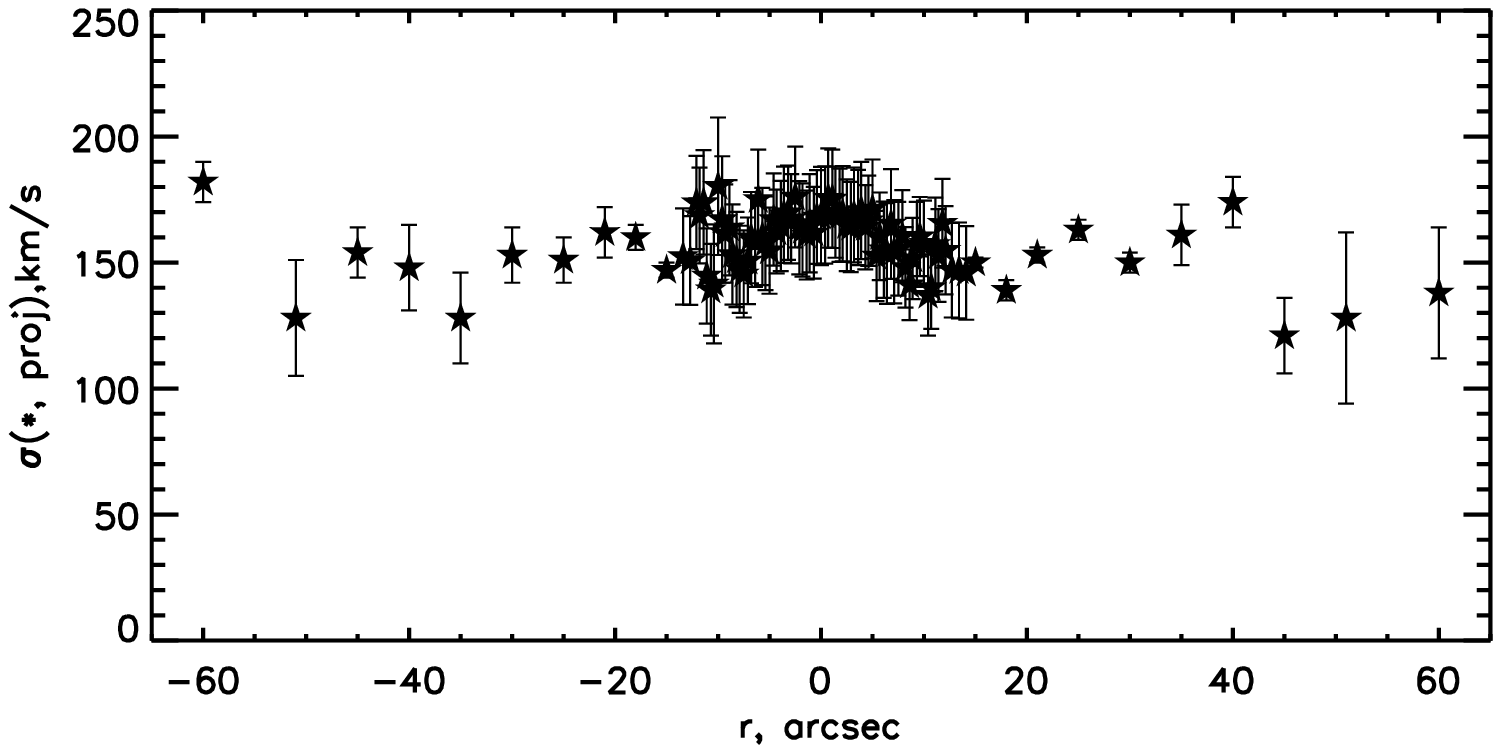}}
\resizebox{\hsize}{!}{\includegraphics{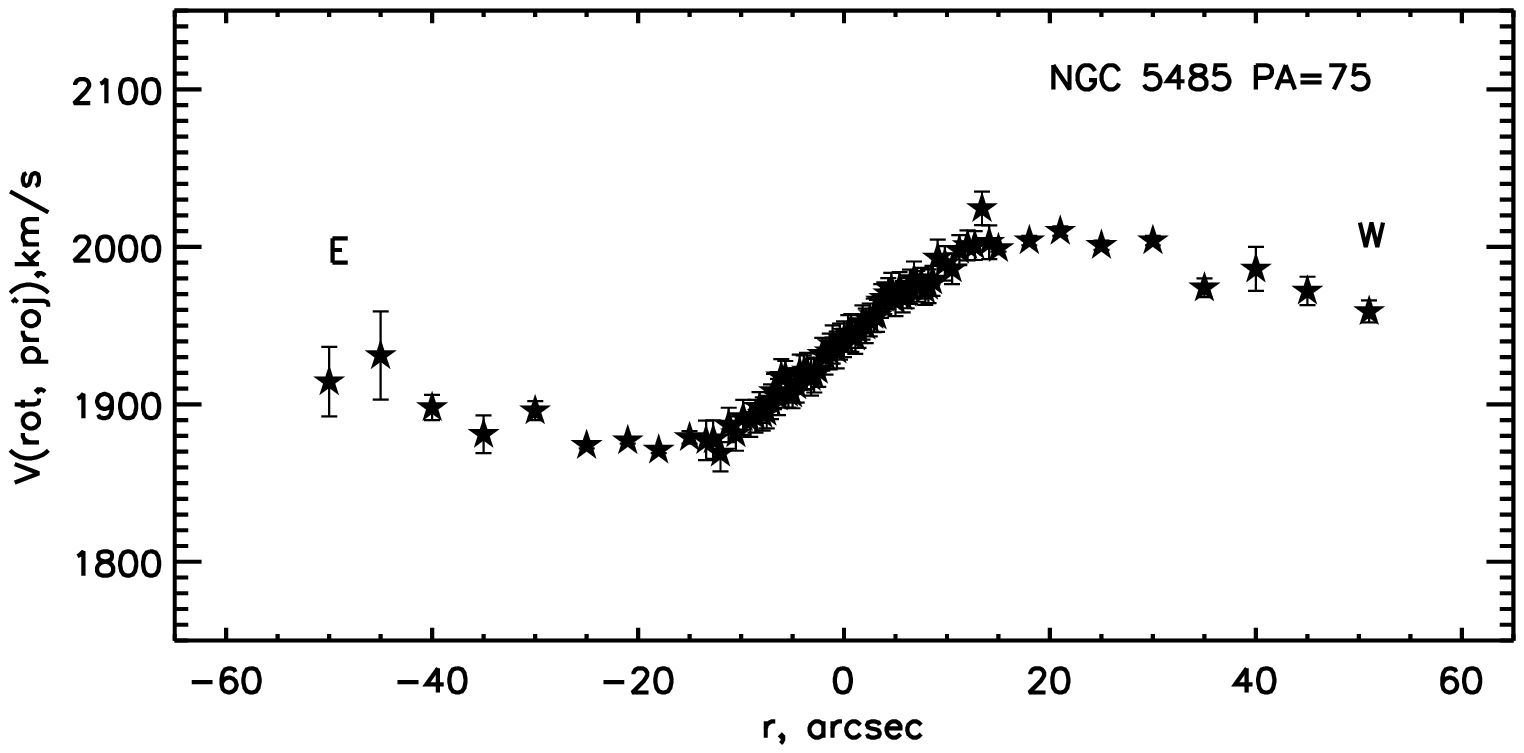}
\includegraphics{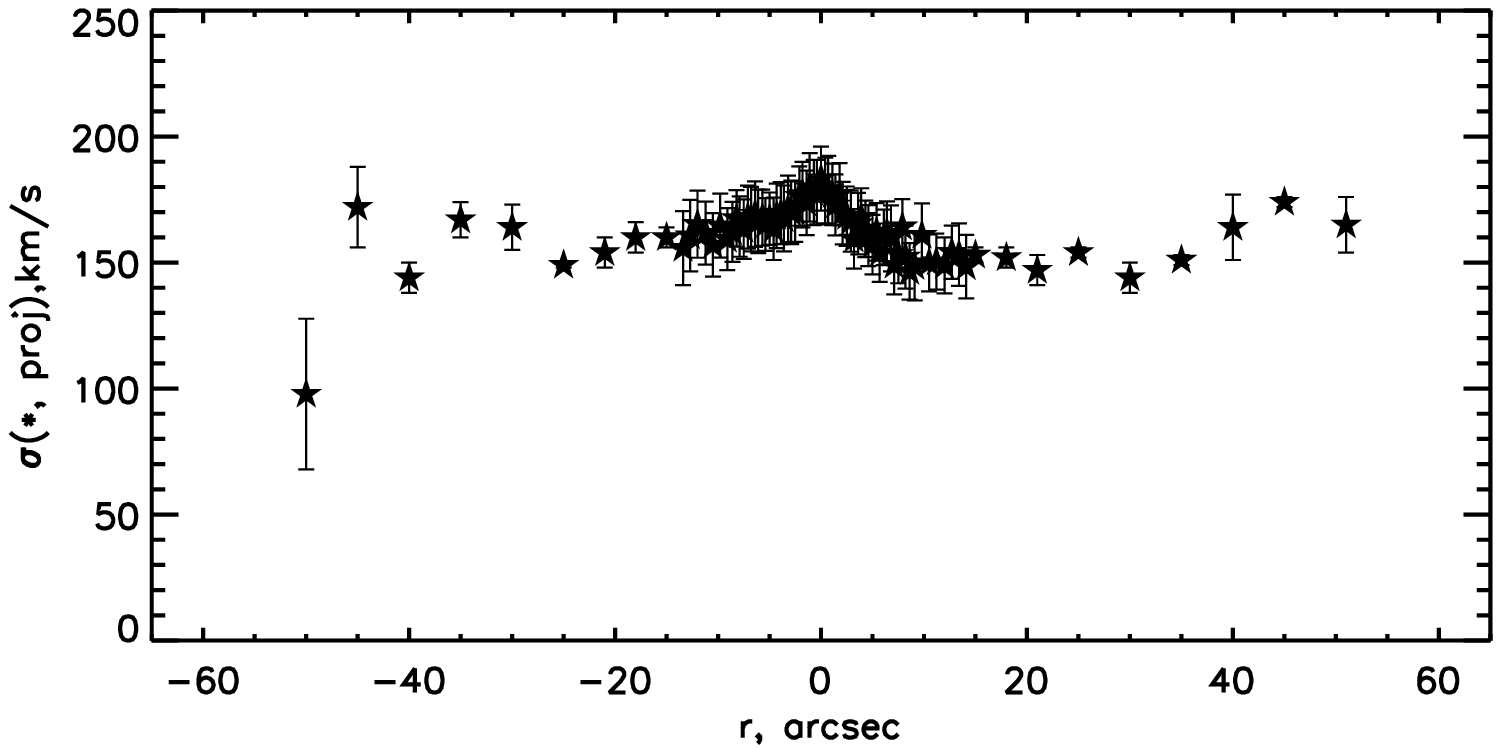}}
\resizebox{\hsize}{!}{\includegraphics{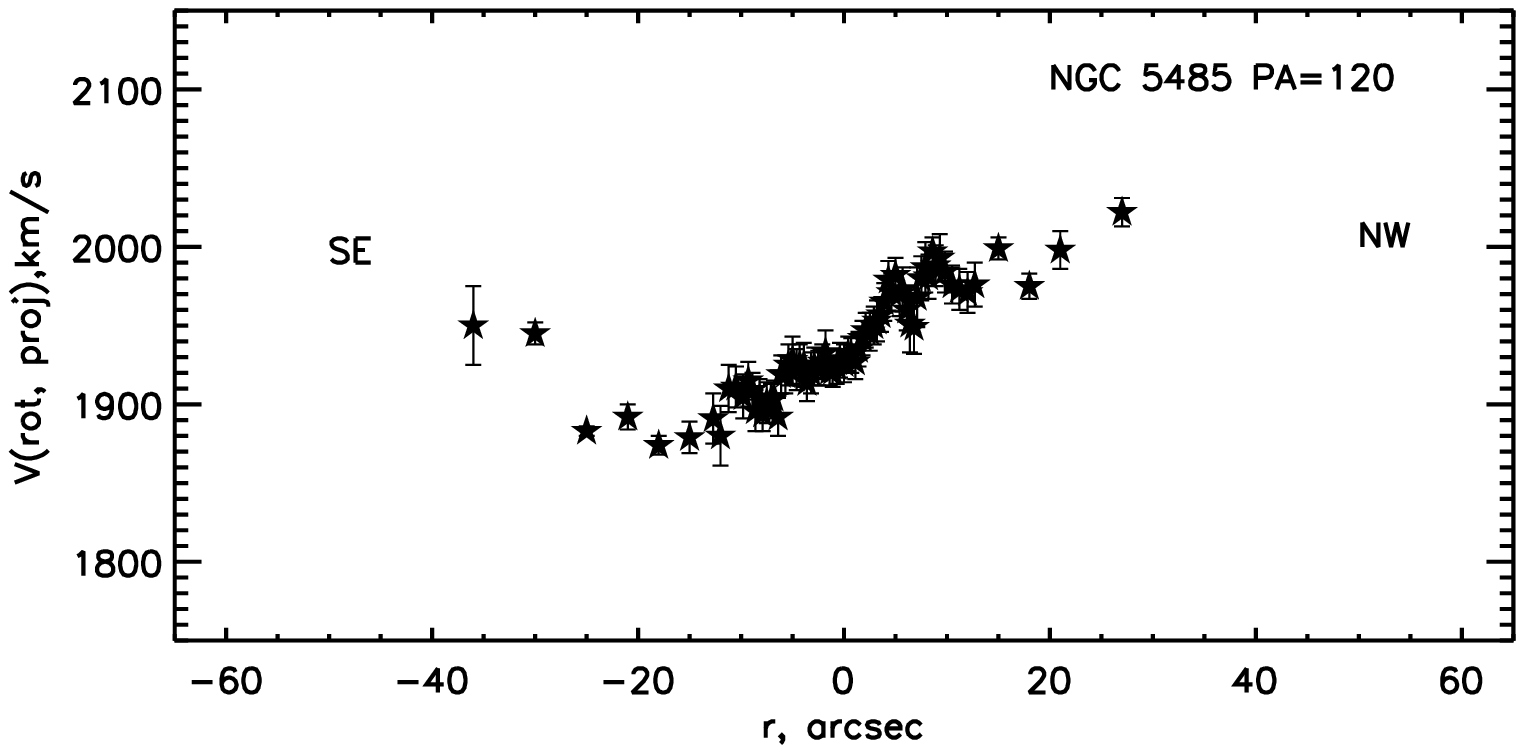}
\includegraphics{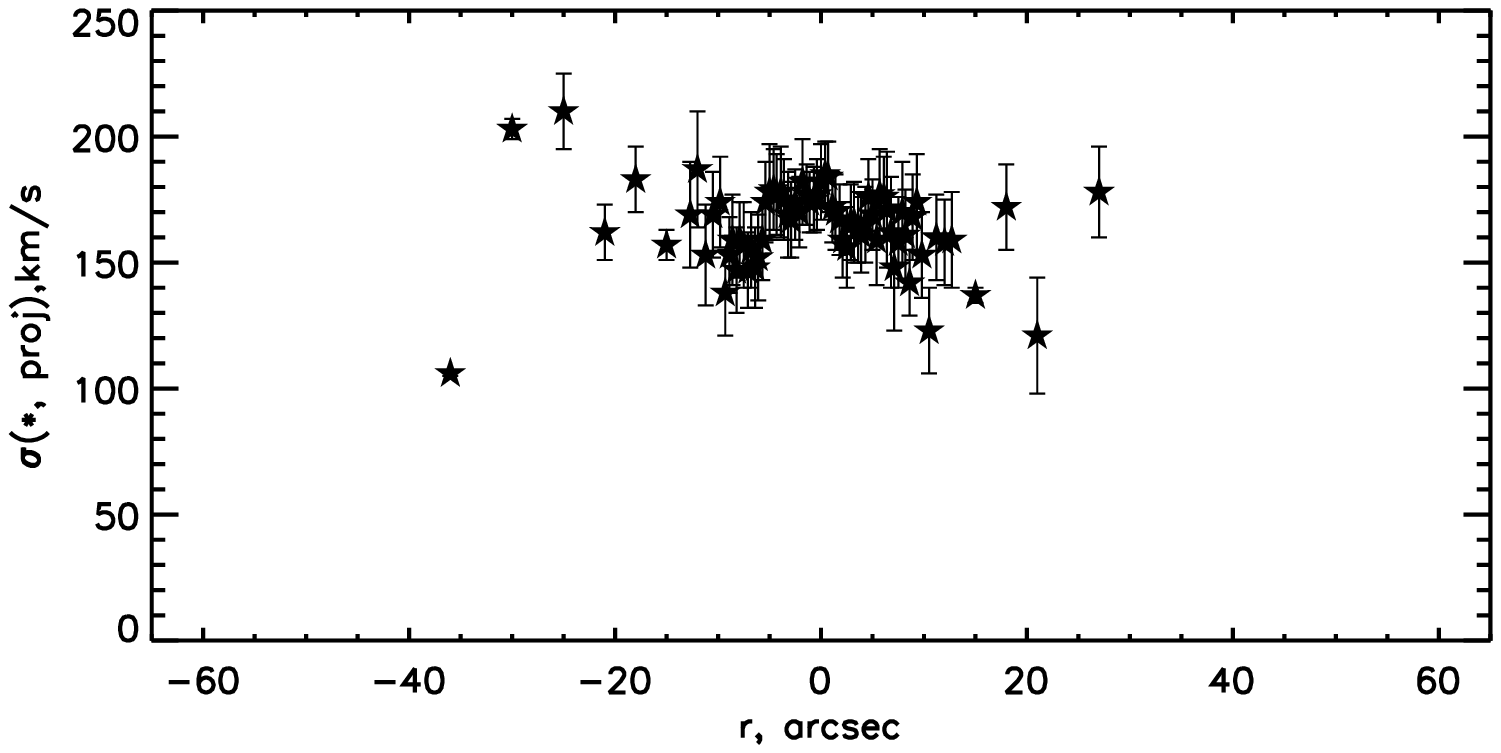}}
\caption{Kinematical cross-sections obtained for NGC~5485 with the SCORPIO spectrograph in the long-slit mode 
at the 6-m SAO RAS telescope at three different position angles: the stellar line-of-sight velocities ({\it left}) 
and the corresponding stellar velocity dispersions ({\it right}).}
\label{n5485longslit}
\end{figure*}

Our long-slit spectroscopy was performed at the 6-m SAO RAS telescope with the SCORPIO multimode
focal reducer \cite{scorpio}. We obtained four cross-sections in total at different position
angles for NGC~502 and several cross-sections at three different position angles for NGC~5485. A full log 
of observations listing the exposures used in our analysis is presented in Table~2. The observations 
were made mostly with a VPHG2300G grism and a slit width of 1 arcsec, that provided a spectral resolution 
of 2.2~\AA\ being sufficient to measure stellar velocity dispersion in the disks. Only the last observation 
of NGC~5485 in 2015 was made with the new SCORPIO-2 version of the instrument \cite{scorpio2}, with
a VPGH1200@540 grism and a spectral resolution of 5~\AA. We measured the Doppler shifts of absorption
lines by cross-correlating the pixel-by-pixel spectra taken along the slit at various distances from the
galactic center with the spectra of bright G8--K3 III stars and with the twilight spectra (G2 class) 
taken on the same nights with the same instrumentation setup. The data turned out to be deep enough 
to allow measuring the stellar kinematics up to the optical boundaries of the disks. The results of our 
measurements, namely, the radial profiles of the stellar line-of-sight velocities and velocity
dispersions, are presented in Fig.~\ref{n502longslit}  and Fig.~\ref{n5485longslit}. 
The radii where the velocity dispersions reached a low-level plateau 
were considered as cross-roads at which and beyond the spectra began to be dominated by the disk light. The
radial ranges of the disk domination derived from the kinematical profiles of NGC~502 and NGC~5485 in this case 
have appeared to be consistent with exponential-shape intervals in the photometric surface brightness profiles 
(for the photometric data, see below). 

\begin{figure}[p]
\includegraphics[width=8cm]{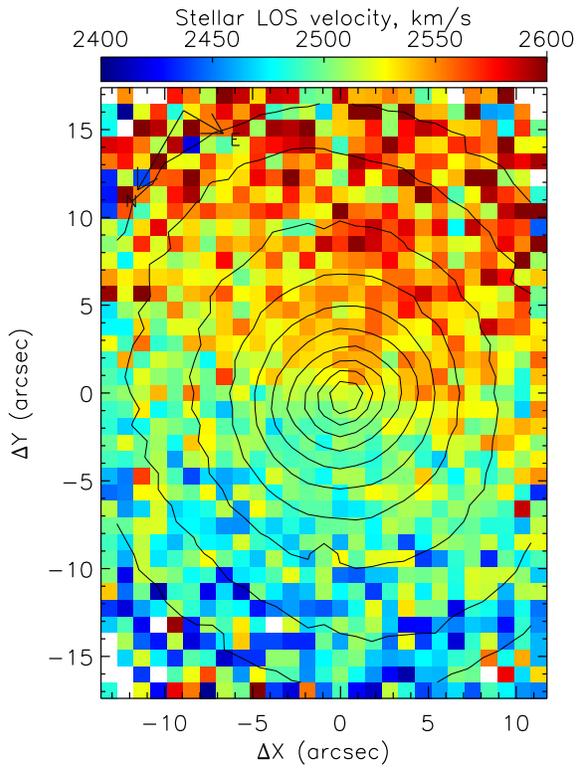}
\caption{Stellar velocity field for NGC~502 calculated using
data from the SAURON integral-field spectrograph.
The orientation of the picture is indicated by the arrows
directed to the north and the east in the upper left corner
of the map. The surface brightness distribution in continuum
at a wavelength of 5100~\AA\ is superimposed by
the isophotes.}
\label{n502field}
\end{figure}

\begin{figure*}[p]
\begin{tabular}{c c}
\includegraphics[width=8cm]{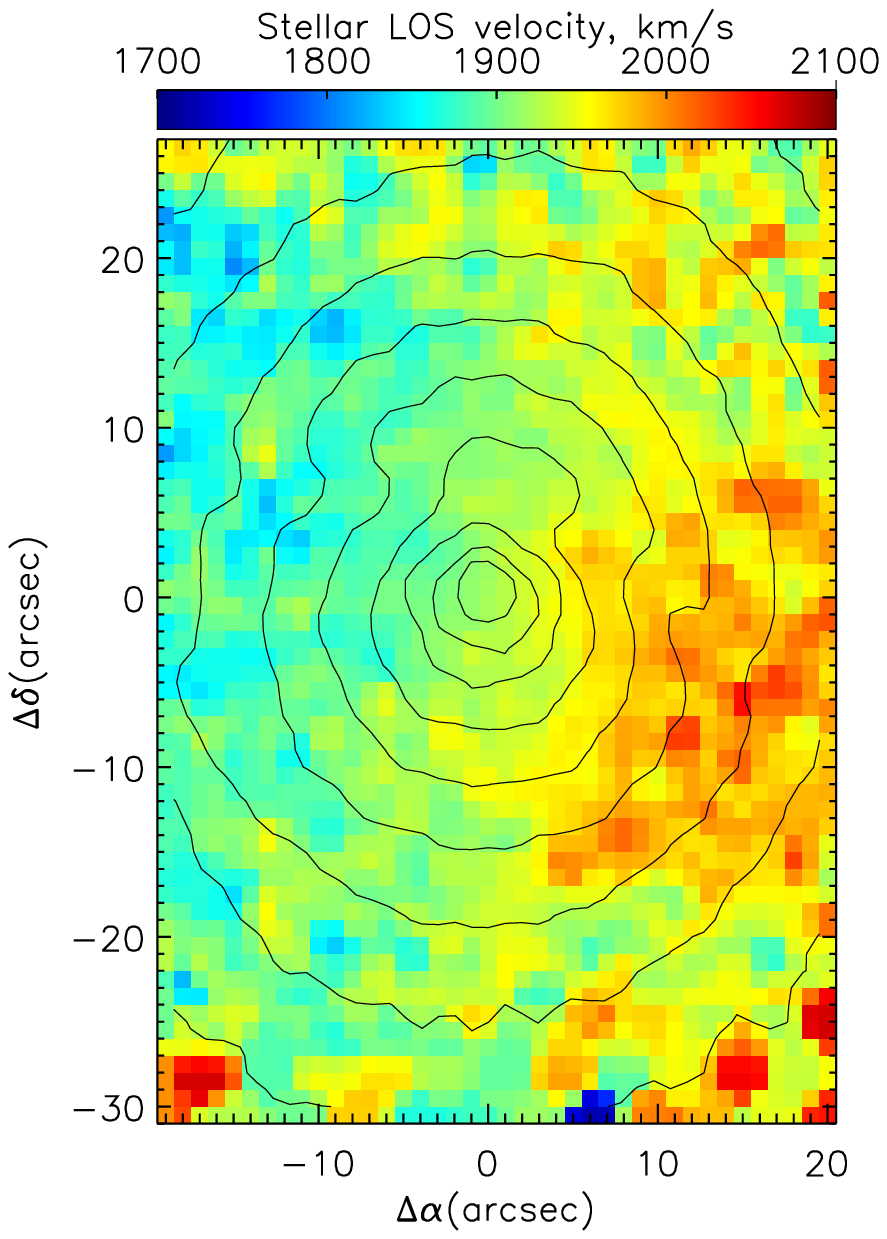} &
\includegraphics[width=8cm]{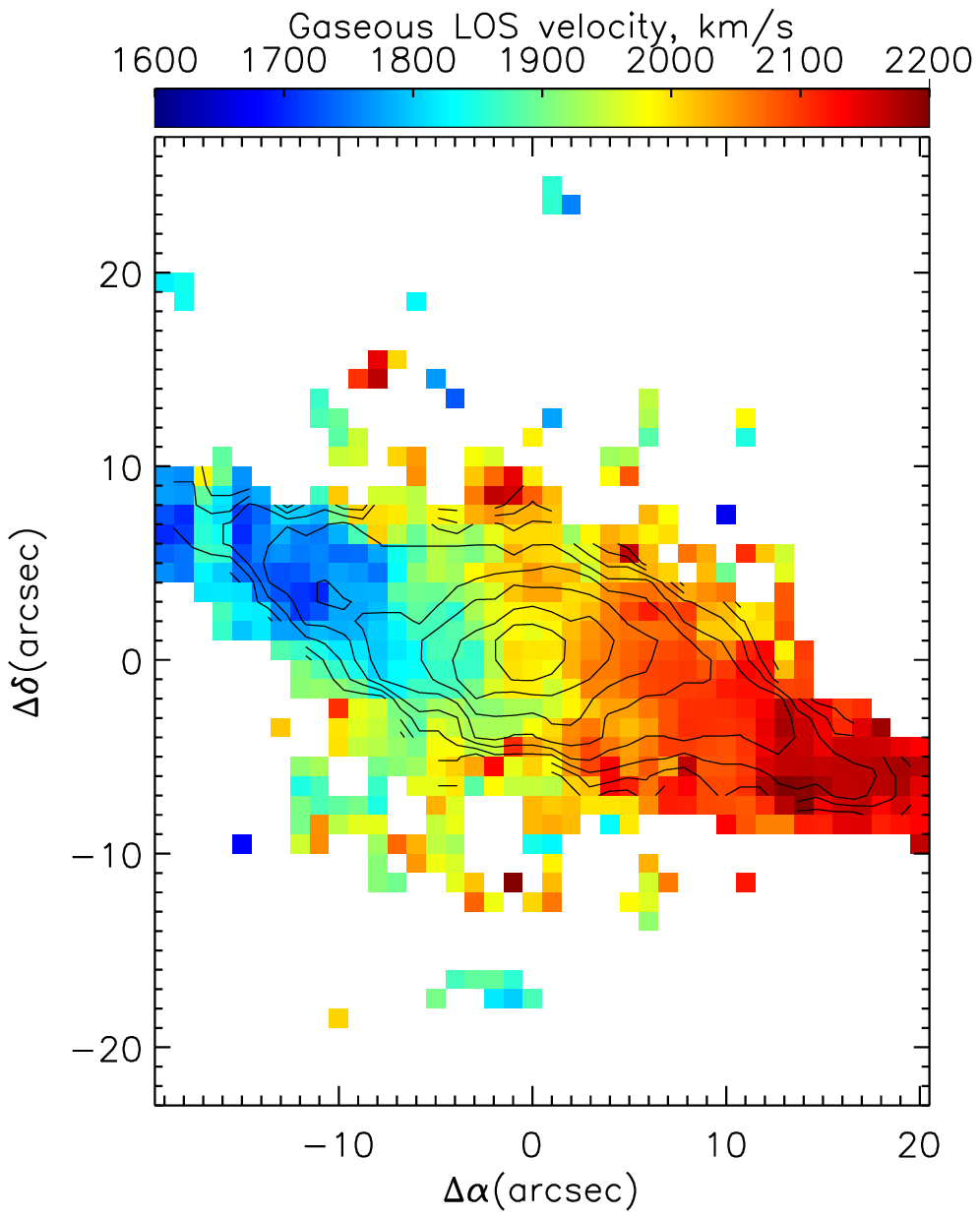}\\
\end{tabular}
\caption{Stellar and ionized-gas (by measuring the [NII]$\lambda$6583 emission line, the strongest 
one in the optical range in this galaxy) velocity fields for NGC~5485 calculated using the data from 
the PMAS integral-field spectrograph obtained as part of the the CALIFA project. The surface brightness 
distribution in continuum at a wavelength of 5100~\AA\ (({\it left}) for the stellar velocity field) 
and the [N II]$\lambda$6583 emission line flux distribution (({\it right}) for the ionized-gas velocity field) 
are superimposed by isophotes.}
\label{n5485field}
\end{figure*}

Apart from the extended kinematical cross-sections obtained
with a long slit, we had some two-dimensional line-of-sight velocity maps for the galactic centers obtained
with integral-field spectrographs at our disposal. Both galaxies were observed within the
ATLAS-3D project \cite{atlas3d_1} with the SAURON spectrograph \cite{sauron} at the
4.2-m William Herschel Telescope at La Palma; the raw data were downloaded from the public ING
(Isaac Newton Group) archive of the Cambridge Institute of Astronomy and were reduced by our original
technique \cite{sil05}. The field of view of the SAURON spectrograph is $33'' \times 41''$, a single spatial
element is $0.94''$, and the spectral resolution is about 4~\AA. In addition, NGC~5485 was also observed as a part
of the CALIFA project \cite{califa_1,califa_3} with the PMAS/PPAK integral-field spectrograph at the 3.5-m telescope 
of the Calar Alto Observatory, and we used the already reduced public data cube with a spectral resolution 
of 500 and a spatial element of $1''$ (the spatial resolution is about $3''$) to construct the velocity fields 
for the stellar component and the ionized gas present at the center of this lenticular galaxy, for
a $60'' \times 40''$ area. The velocity fields for NGC~502 and NGC~5485 are shown 
in Fig.~\ref{n502field} and Fig.~\ref{n5485field} . 
They were analyzed by the tilted-ring method under Moiseev's modification -- by the DETKA code \cite{moiseev04}.
The orientation of the kinematical major axis, which in the case of a circular rotation must be coincident with 
the line of nodes of the disk plane, was traced up to a distance from the center of about $20''$ in NGC~502 
and up to $25''$ in NGC~5485. Although these distances are defined by the quality of the stellar velocity fields, 
by chance, they are at the beginning of the galactic exponential disks. 

The photometric analysis of the  
structure of these galaxies is abundantly presented in the literature. In particular, deep photometry of NGC~502 
performed with the SCORPIO focal reducer of the 6-m SAO RAS telescope in the direct-image mode was
described in detail by Il'ina and Sil'chenko \cite{n524phot}; the presence of wide stellar rings (surface brightness
excesses) between $8'' - 16''$ and $35'' - 45''$ was pointed out. A two-dimensional decomposition of
the NGC~5485 image was undertaken by M\'endez-Abreu et al. \cite{mendez_abreu}, Laurikainen et al. \cite{lauri10,nirs0s},
and Guti\'errez et al. \cite{erwin11}, where significantly differing metric parameters of the galactic disk
and bulge are presented. We additionally undertaken an isophotal analysis of the $r$-band images of both
galaxies based on SDSS data, release 9 \cite{dr9}, and of the 4.5$-\mu$m image of NGC~5485 retrieved from the
public database of the S4G survey \cite{s4g}, though the latter image is at the edge of the field of view 
of the Spitzer telescope and, therefore, the characteristics of the outer isophotes are not very sure. 
Figure~5 compares the kinematical and photometric parameters: the ellipticities measured within our
isophotal analysis are compared with $1-\cos i$, where the inclinations of the disk plane to the plane of the
sky, $i$, were obtained by the tilted-ring method applied to the two-dimensional velocity fields; the orientations
of the kinematical and photometric major axes in the central regions of the galaxies are also compared. As
it was expected, the kinematical and photometric major axes in the central regions of the galaxies have
appeared to be strongly misaligned. The parameters of the isophotes farther from the center, into the region 
where the exponential stellar disks dominate in the surface brightness, are also traced in Fig.~\ref{iso}.

\section{THE ELLIPTICAL OUTER DISKS IN NGC 502 AND NGC 5485}

Let us consider in detail orientations of the NGC~502' and NGC~5485' disks in
space by using all the available photometric and kinematical information.

\begin{figure*}[p]
\rule{0pt}{51pt}
\vspace*{1cm}
\resizebox{\hsize}{!}{\includegraphics{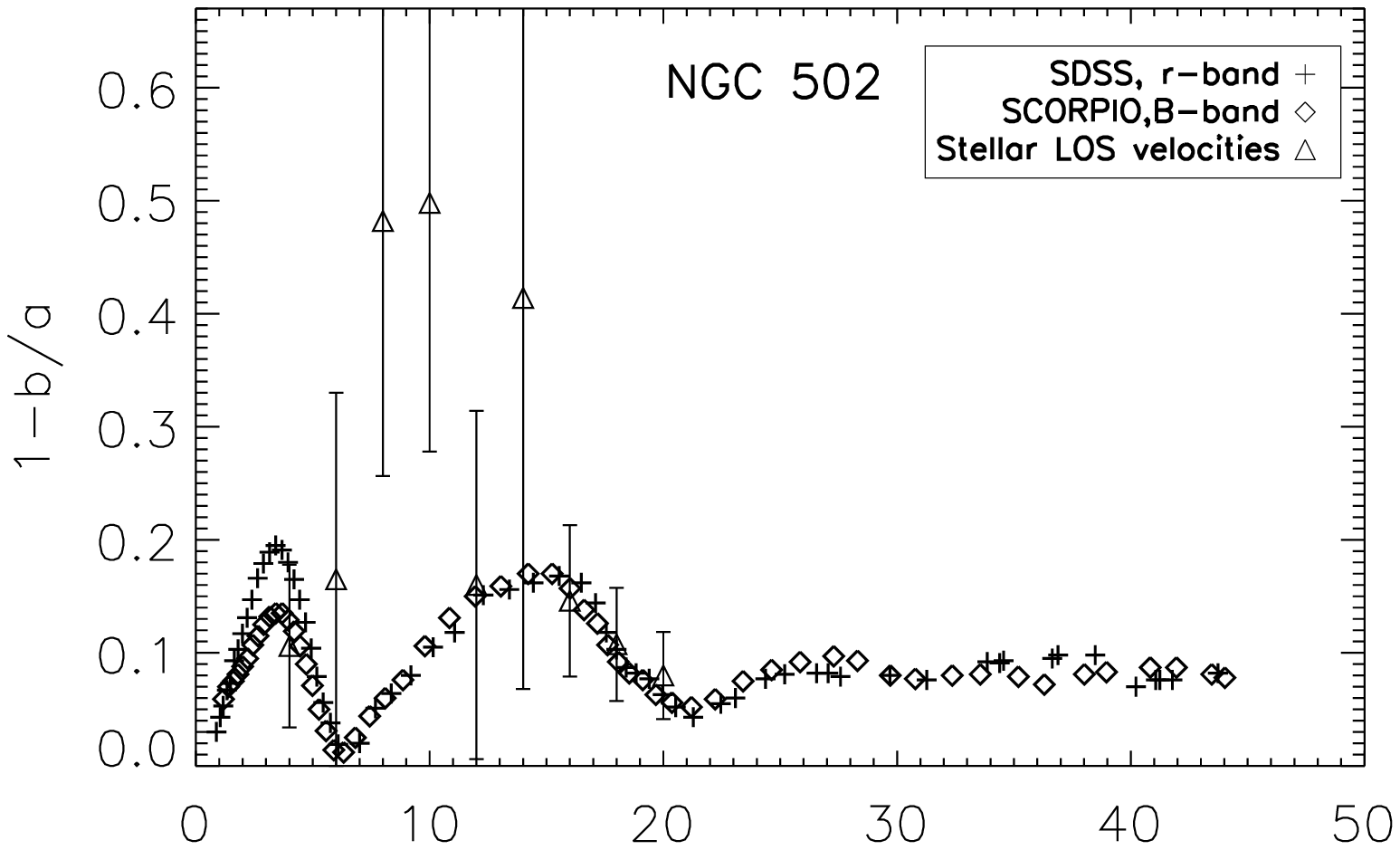} 
\includegraphics{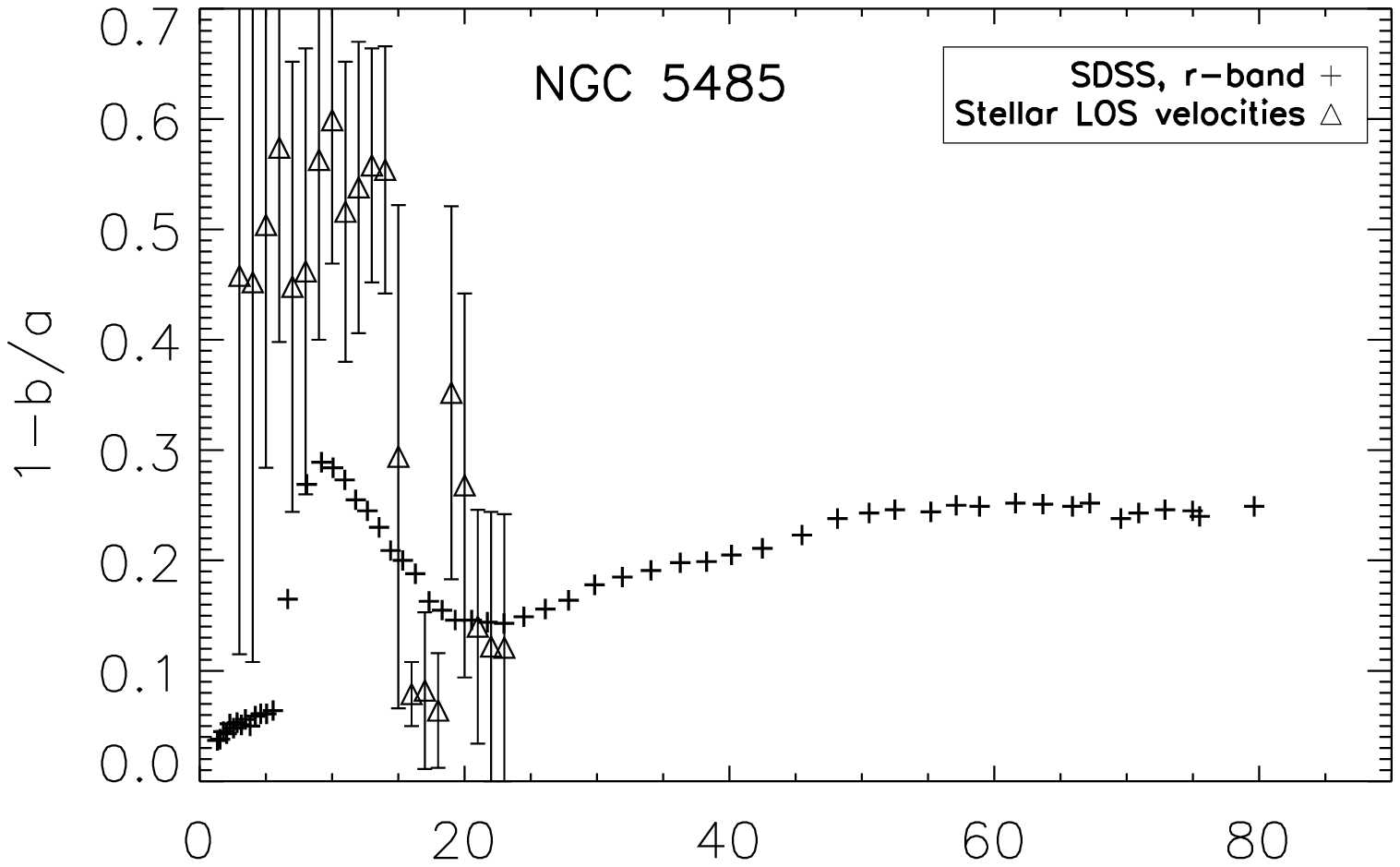}}
\vspace*{1cm}
\resizebox{\hsize}{!}{\includegraphics{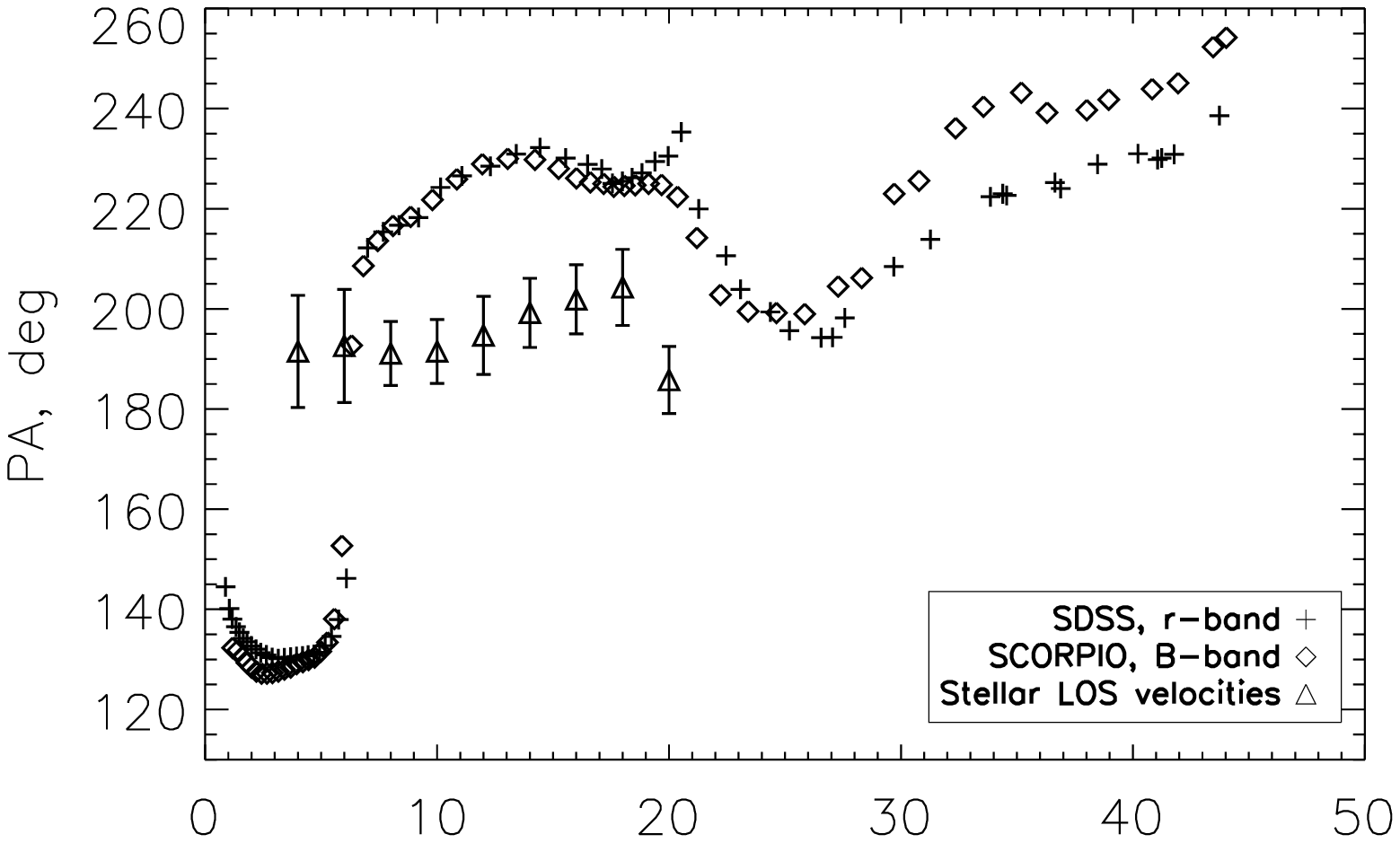} 
\includegraphics{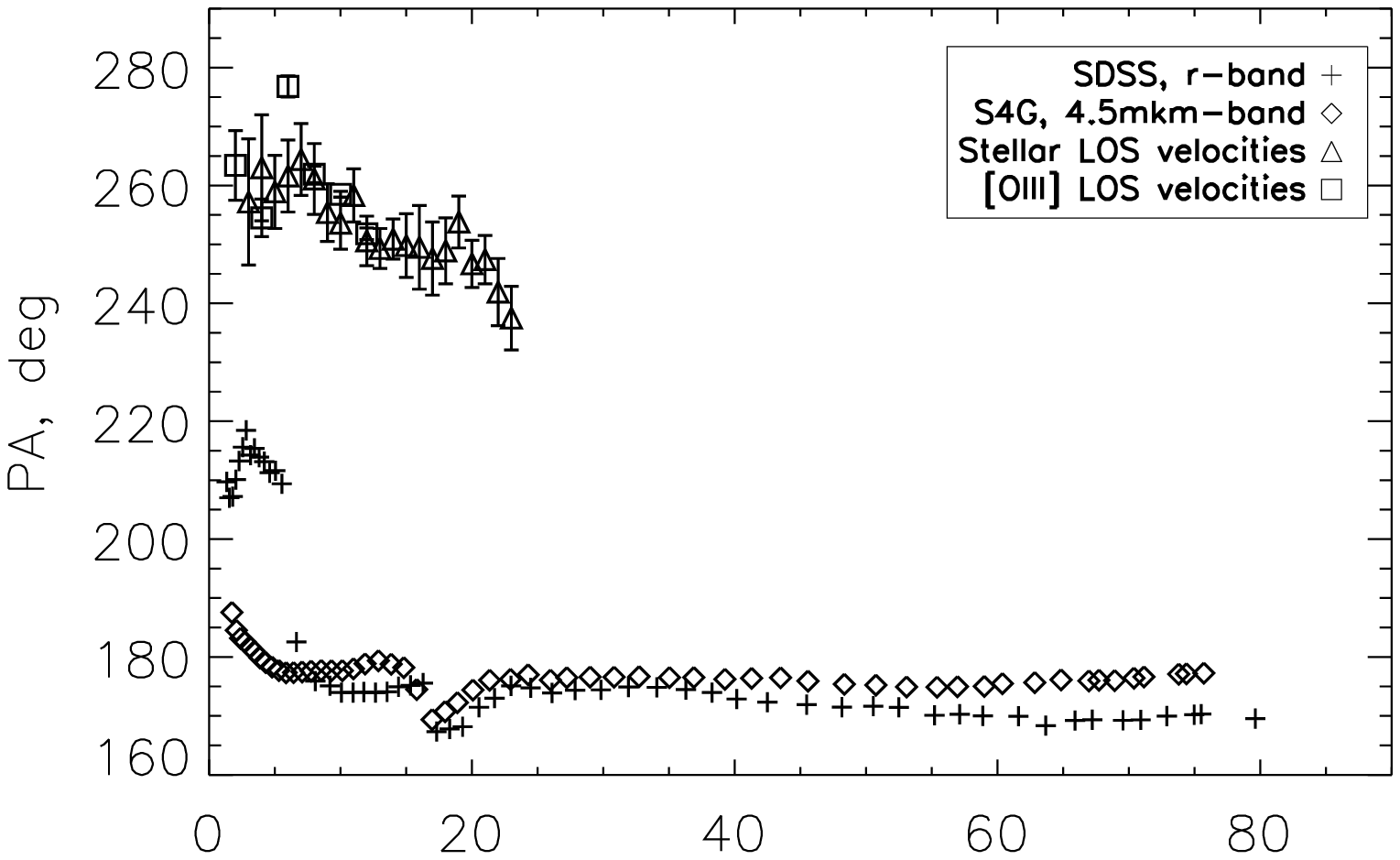}}
\caption{Isophotal ellipticities and orientations of the photometric and kinematical major axes 
in NGC~502 ({\it left}) and NGC~5485 ({\it right}).}
\label{iso}
\end{figure*}

\begin{figure*}[p]
\begin{tabular}{c c}
\includegraphics[width=8cm]{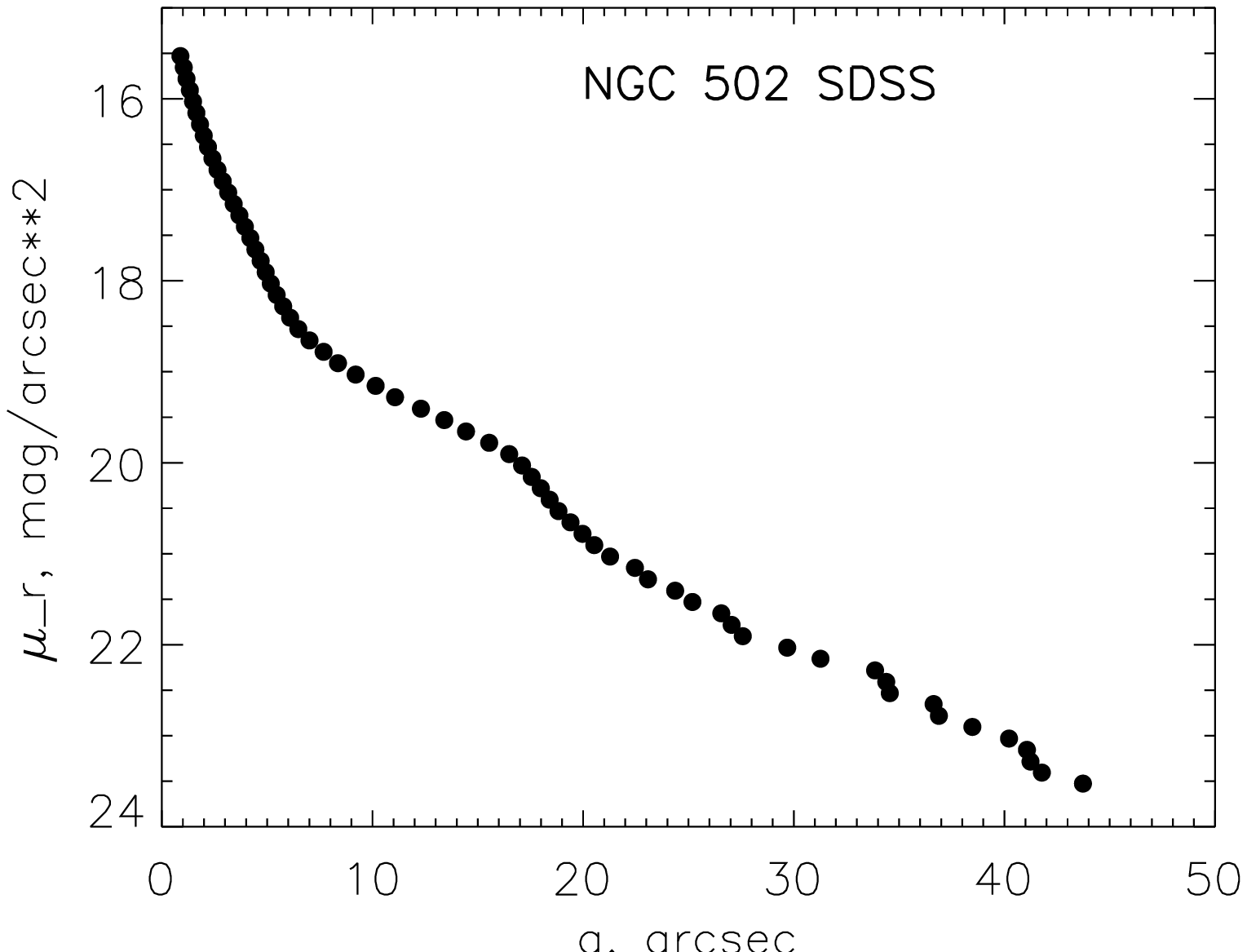} &
\includegraphics[width=8cm]{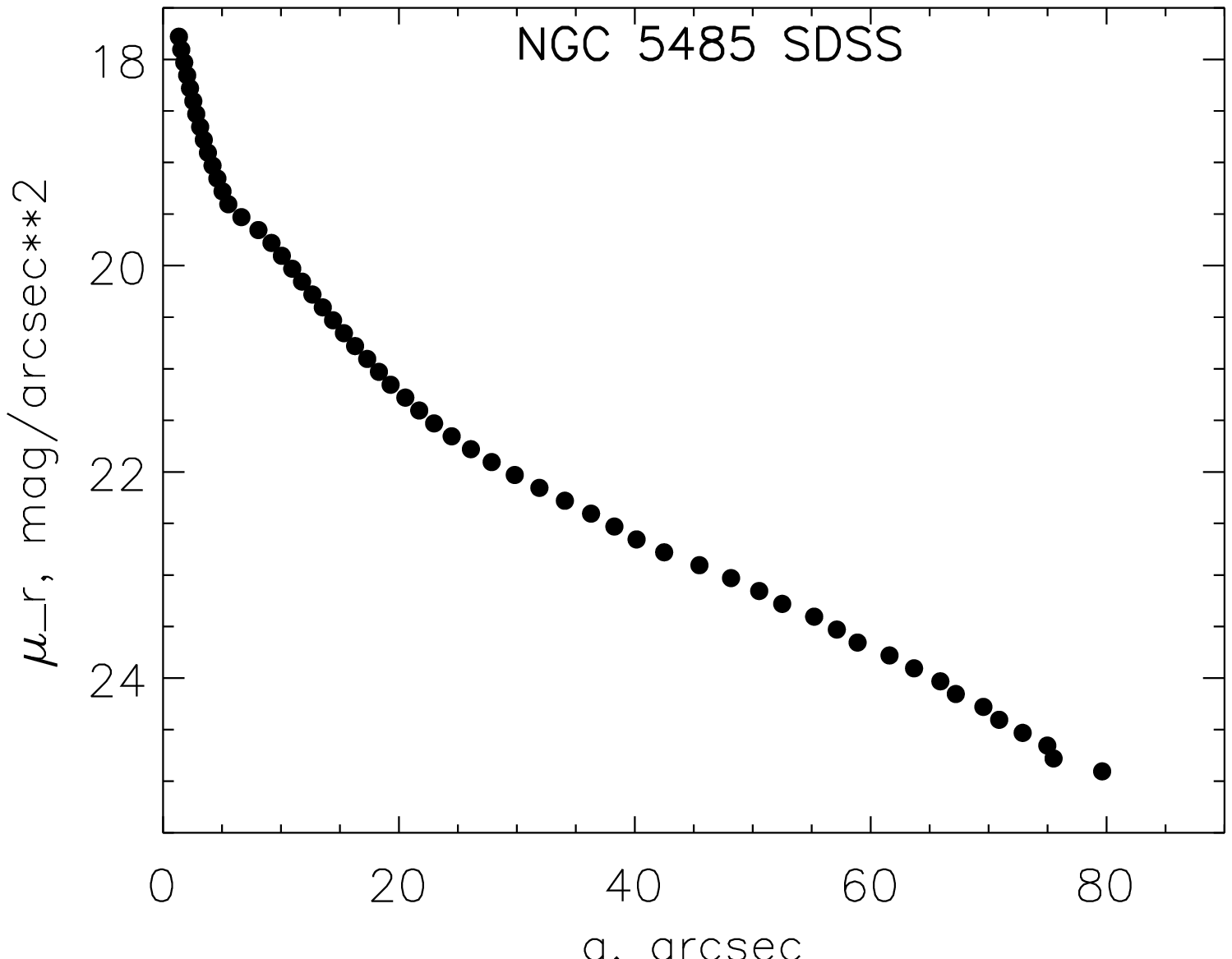}\\
\end{tabular}
\caption{Azimuthally averaged (with isophotal parameters sliding along the radius) surface brightness profiles 
for NGC~502 ({\it left}) and NGC~5485 ({\it right}), from the SDSS data.}
\label{prof}
\end{figure*}

\begin{figure*}[p]
\begin{tabular}{c c}
\includegraphics[width=8cm]{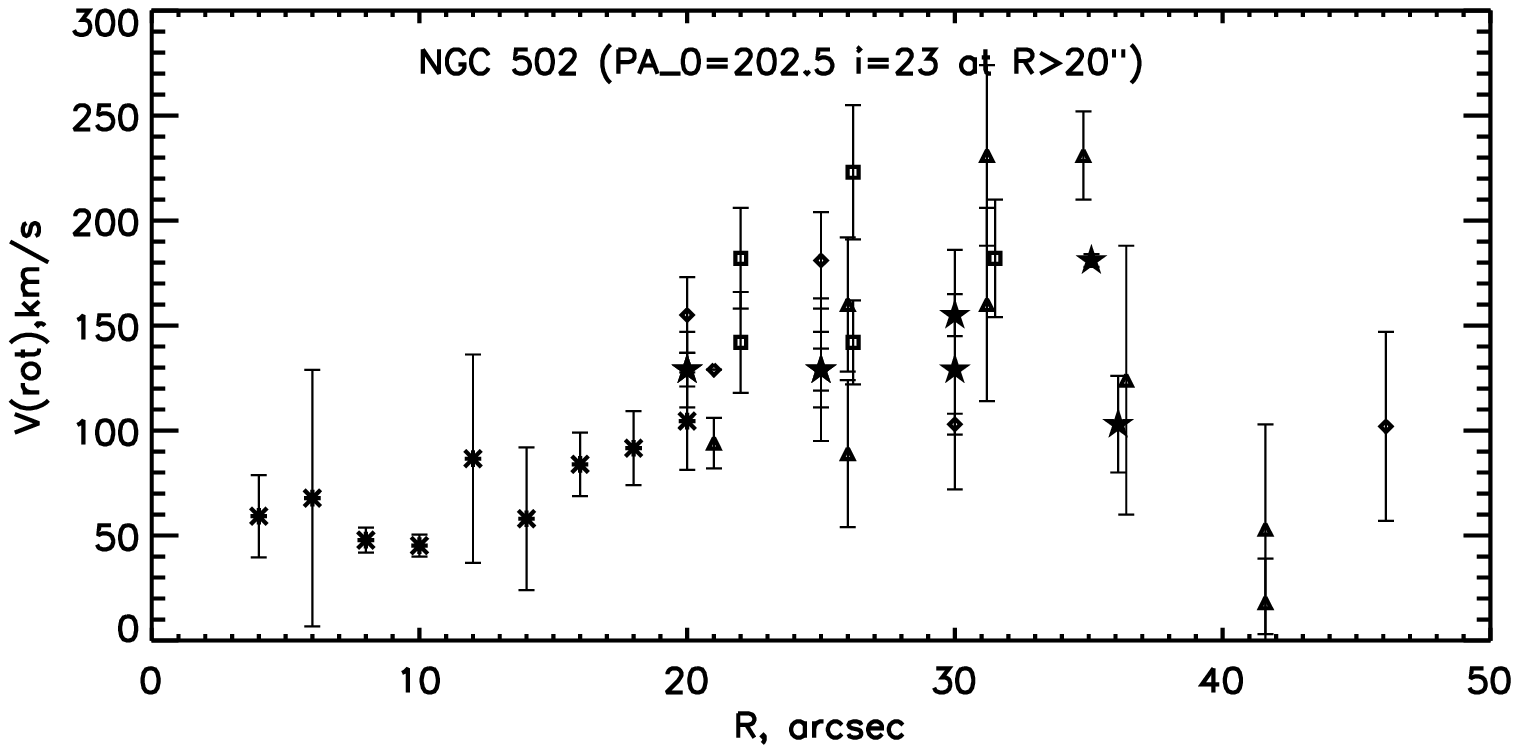} &
\includegraphics[width=8cm]{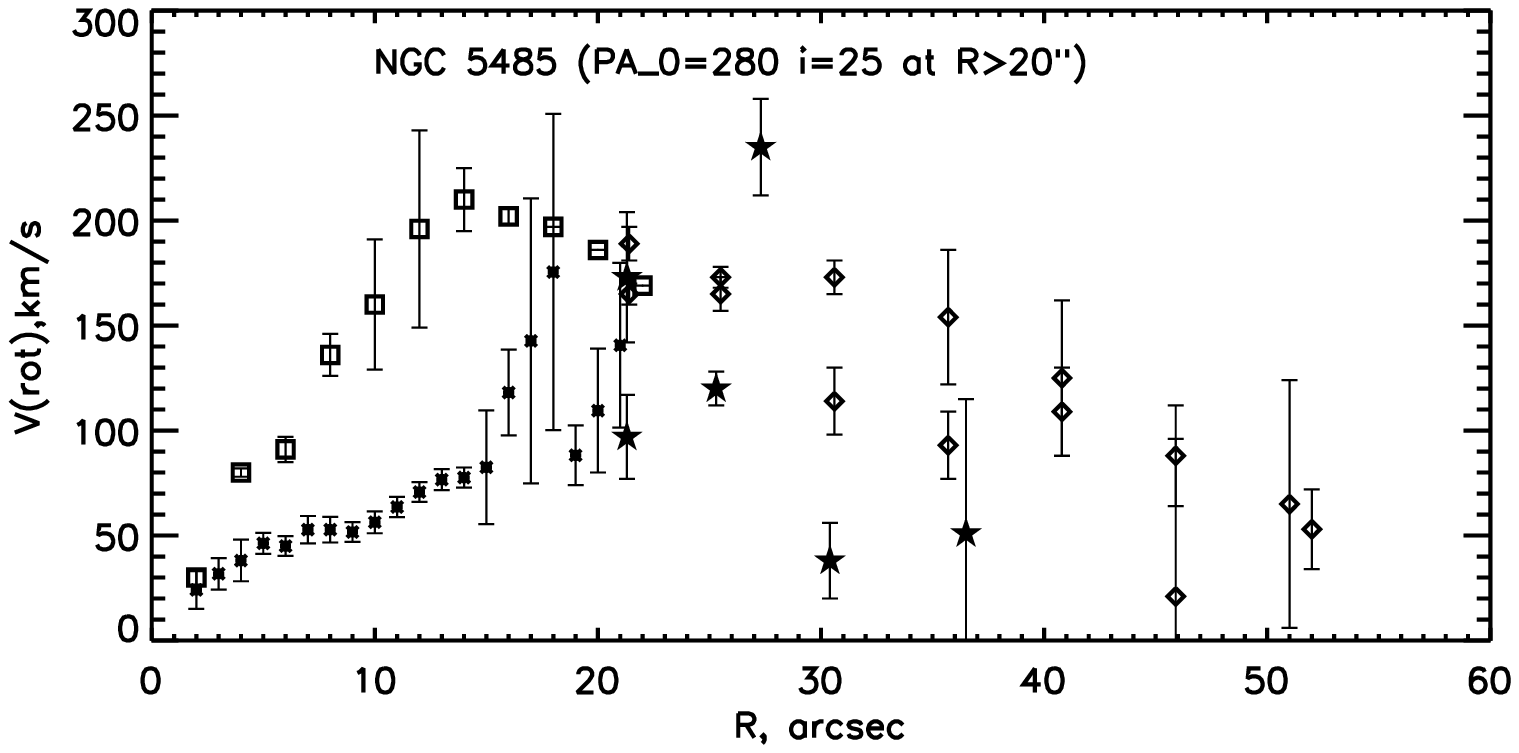}\\
\end{tabular}
\caption{Radial circular velocity profiles under the assumption of a circular rotation of the stellar disks 
in NGC~502 ({\it left}) and NGC~5485 ({\it right}) (uncorrected for the asymmetric drift). The inner regions, 
$R < 20''$, are the stellar rotation curves (asteriks) constructed by the tilted-ring method from the two-dimensional 
line-of-sight velocity maps. In the outer regions, various other symbols indicate the stellar line-of-sight 
velocity profiles at various position angles recalculated into the circular rotation
velocities with the orientation parameters of the rotation planes specified at the top of the plots. 
For NGC~5485, the squares also indicate the gas rotation curve obtained through a one-dimensional cut along the kinematical 
major axis of the line-of-sight velocity field for a gaseous disk seen nearly edge-on.}
\label{vrot}
\end{figure*}

\begin{figure}[p]
\includegraphics[width=8cm]{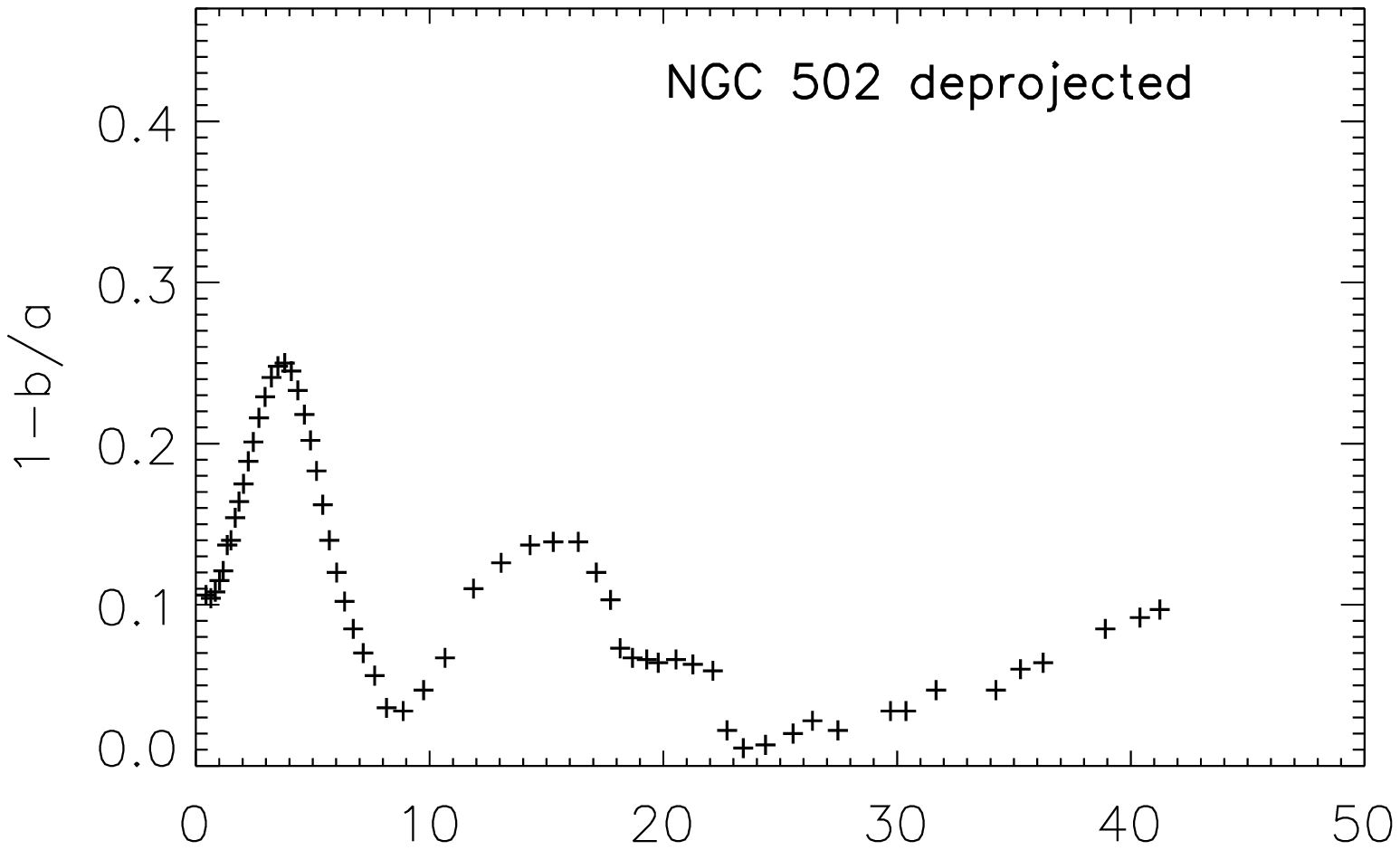}
\caption{Intrinsic ellipticities of the isophotes in the image of NGC~502 (SDSS, the $r$-band) deprojected by assuming
the following orientation of the galactic plane in space:
the disk inclination to the plane of the sky is $i = 23^{\circ}$, the position angle of the disk line of nodes is
$PA_0 = 202.5^{\circ}$.}
\label{depro}
\end{figure}

\noindent
{\bf NGC 502.} According to the results of our isophotal analysis, after inspecting the surface brightness 
profile (\cite{n524phot}; also see Fig.~\ref{prof}, left), we conclude that the inner ring in NGC~502 is localized 
near the radius of $15''$; it is also clearly seen by eye in the combined-coloured SDSS image.
Since we have also the results of integral-field spectroscopy, namely, the two-dimensional
stellar velocity field just for this radial zone, we find that the kinematical and photometric major
axes in this region are misaligned by some $30^{\circ}$. Obviously, the inner ring of NGC~502 has elliptical shape. 
Starting from the radius of $20''$, the surface brightness profile reveals exponential shape, and the stellar velocity dispersion profile (Fig.~\ref{n502longslit}) flattens out at $\sigma _* \sim 60$ km s$^{-1}$. Obviously,
here we deal with the area of stellar disk domination in the surface brightness and in the matter density. 
What is the orientation of this disk in space? According to the integral-field spectroscopy, the kinematical 
major axis reaches $PA \approx 203^{\circ}$ at the radius of $R = 16'' -18''$; the isophotes in
the radius range of $R = 22'' -30''$, beyond the influence of the inner ring, in the region of exponential disk
domination, show approximately the same position angle of the major axis: $PA_{phot} = 204^{\circ}$ according
to our B-band data from SCORPIO \cite{n524phot} and $PA_{phot} = 201^{\circ}$ according to the $r$-band SDSS data. 
We would conclude that we see a round stellar disk oriented nearly face-on in the
radius range of $R = 22'' -30''$: the apparent ellipticity of the isophotes in this radius range is $1 - b/a = 0.08$,
that corresponds to the inclination of $23^{\circ}$ for the case of a very thin disk and is consistent with the 
photometric inclination of $24^{\circ}$ from the HYPERLEDA database (see Table~1). Interestingly, 
the kinematical inclination at $R > 16''$ is also consistent with the photometric one under the assumption 
of a thin disk (Fig.~\ref{iso}, left). We shall take these orientation angles, $i = 23^{\circ}$ and the position angle
of the line of nodes $PA_0 = 202.5^{\circ}$, as characterizing the orientation of the NGC~502' disk in space, 
and shall attempt to bring all the four kinematic cross-sections of Fig.~\ref{n502longslit} into one
rotation curve under the assumption of circular bulk motions. The result of our calculation
is shown in Fig.~\ref{vrot},left. Inside the radius of disk-domination starting, at $R < 20''$, 
the rotation curve is shown which is constructed from the SAURON two-dimensional velocity field by 
the method of tilted, intrinsically circular rings fixing the inclination of the rotation plane and 
allowing the position angle of the line of nodes to change freely along the radius; whereas at $R > 20''$ 
different symbols indicate our four cross-sections of Fig.~\ref{n502longslit} with a long-slit velocity 
measurements recalculated to the circular rotation velocities under the fixed orientation angles of
the rotation plane specified above. The result has turned out to be good: within the error limits, in the range
$R = 22'' -30''$, all four cross-sections have shown similar maximum stellar rotation velocities, 
about 145 km s$^{-1}$, which, besides all, also corresponds to the luminosity of NGC~502 if we refer 
to the Tully–Fisher relation, for example, from Theureau et al. \cite{tf_theureau}. However, beyond the
standard optical radius of the stellar disk, at $R > 35''$, the calculated rotation velocity drops 
sharply (Fig.~\ref{vrot},left), the position angle of the isophotes increases sharply (Fig.~\ref{iso}, left bottom), 
and another ring is observed in the surface brightness profile \cite{n524phot}. Obviously, the outer wide ring
of NGC~502 is also elliptical. Figure~\ref{depro} shows the isophotal ellipticity for the image of NGC~502 deprojected
toward the view face-on by using the orientation parameters specified above. We see two distinct ellipticity peaks 
in the inner region, at $R = 3.5''$ and $R=15''$, corresponding to the nuclear bar and to the inner ring, 
while in the outer parts of the disk, at $R > 30''$, the intrinsic disk ellipticity begins to rise monotonically, 
and reaches 10\%\ at $R = 42''$. The galaxy's deep image obtained with the MegaCam camera at the 
Canada–France–Hawaii Telescope (CFHT) shows numerous stellar shells in the outer regions \cite{atlas3d_29}. 
Obviously, despite the relative isolation of NGC~502, it suffered a number of minor mergers, probably with 
dwarf galaxies devoid of gas, which could produce an oval distortion of some zones at fixed radii 
of the (globally) dynamically cold stellar disk.

\noindent
{\bf NGC 5485.} As it can be seen in Fig.~\ref{iso}, right bottom, the kinematical major axis in the central part 
of NGC~5485 is exactly perpendicular to the photometric major axis. This allowed Krajnovic et al. \cite{atlas3d_2} to put
NGC~5485 in the rare subclass of galaxies (consisting of only two objects in the nearby, $D < 42$~Mpc, Universe) 
with the so-called prolate rotation, when the stellar rotation axis coincides with the longest axis of the spheroid 
in a triaxial potential. Even significantly earlier than the ATLAS-3D survey, Wagner et al. \cite{wagner88} stated 
prolate rotation in NGC~5485 by analyzing long-slit spectroscopy data up to a distance of $22''$ from the center. 
Actually, no one of them determined an exact orientation of the stellar rotation axis in this galaxy. These authors
only compared the position angles of the kinematical and photometric major axes, mainly for the central
region of the galaxy, where the bulge dominates in the surface brightness. The idea of general triaxiality
of NGC~5485, when the entire galaxy is an ellipsoid with three different axes, without a flat disk, comes into conflict 
with the constancy (along the radius) of the position angle of the photometric major axis (isophotes) over the entire 
galaxy and with the constancy of the apparent isophotal ellipticity within the area dominated by the exponential 
galactic disk at $R > 30''$. All the photometrists \cite{nirs0s,erwin11} classified this galaxy as unbarred lenticular 
one, i.e., as an axisymmetric body. Since there is a dust lane near the galactic center and since the gaseous disk
coincident with it rotates with the same orientation of the kinematical major axis as the stars (Fig.~\ref{n5485field}), 
it is tempting to suggest that we see rotation of the stellar component formed from accreted gas with
a polar angular momentum at the galactic center up to a distance of $22''$ from the nucleus. However,
whereas the gas is concentrated in a disk seen nearly edge-on, with an inclination of about $70^{\circ}–80^{circ}$ 
in its outer parts, the rotation of the stellar component analyzed with the tilted-ring method provides evidence
for a much more moderate inclination of the rotation plane ($45^{\circ}$ according to the SAURON data and $\le 37^{\circ}$
according to the CALIFA data); and, obviously, there are no signatures of a strongly inclined ``nested'' stellar disk  
in the isophotes. Even greater difficulties arise with the interpretation of the stellar rotation beyond the central part 
of the galaxy and beyond the polar gas concentration. If we attempt to bring the three cross-sections 
in Fig.~\ref{n5485longslit} into one circular rotation curve under the assumption of an axisymmetric disk using
the photometric orientation angles, $PA_0 = 170^{\circ}$ and the inclination of $45^{\circ}$, then recalculating 
the cross-section taken at $PA = 75^{\circ}$ would give a rotation velocity greater than 1500 km s$^{-1}$, while recalculating the cross-section taken at $PA = -14^{\circ}$ would give rotation velocities close to zero. 
Obviously, this approach does not work. Formally, we can attempt to select the orientation of the plane of 
quasi-circular stellar rotation basing on the similarity of the spectral cross-sections at $PA = 75^{\circ}$
and $120^{\circ}$ (Fig.~\ref{n5485longslit}); the kinematical major axis must be somewhere between these directions.
Figure~\ref{vrot}, right, presents the result of successful convergence of the three kinematical cross-sections 
into one rotation curve made for the following orientation of the plane of quasi-circular rotation: $PA_0 = 100^{\circ}$ 
and $i = 25^{\circ}$. Under such inclination of the rotation plane, the maximum disk rotation velocity turns out to
be about 180 km s$^{-1}$, which agrees excellently with the Tully–Fisher relation in the near infrared by
Theureau et al. \cite{tf_theureau}. The inner part of the circular rotation curve calculated 
from the gaseous velocity field by applying one-dimensional digital cut along the major axis to the ionized-gas 
line-of-sight velocity field presented in Fig.~\ref{n5485field}, right, also joins excellently to (or is continued by) 
this rotation curve of the stellar component. However, such orientation angles have no analog whatsoever in the 
photometric characteristics of the galaxy. And at radii greater than $40''$, the rotation curve in Fig.~\ref{vrot}, right,
 falls to zero, that completely disagrees with the expected shape of the rotation curve for an exponential disk with a
scalelength of 4.5~kpc, or about $33''$ \cite{lauri10} that must reach its maximum at a radius approximately equal 
to two exponential scalelengths. The hypothesis of axial symmetry, even of the partial one, in the case of NGC~5485 
should apparently be abandoned.

\section{CONCLUSIONS}

We have investigated in detail two unbarred lenticular galaxies for which our previous spectroscopic
data have shown a misalignment of the kinematical and photometric major axes. The anomaly of the large-scale
stellar kinematics in NGC~502 has turned out to be related to the presence of two wide elliptical stellar rings 
that border an otherwise normal stellar disk from the inside (adjacent to the bulge) and from the outside
(near the optical boundaries of the galaxy). 

As for NGC~5485, the situation is obviously more complicated: 
we have failed to identify an axisymmetric structural component in the galaxy at any distance from
the center. The rotation axis of the galactic bulge coincides in projection with the major axis of the
isophotes; this component may be genetically related to the accreted polar gas observed at the center of 
NGC~5485 as a disk consisting of gas and dust seen nearly edge-on. However, the central stellar component
of NGC~5485 is definitely not a thin disk but most likely a bulge, i.e., the so-called prolate spheroid. It should
be noted that the stellar population of the bulge is old, $T_{SSP} \sim 9$~Gyr \cite{silsaus0}. It means that even 
if this component is formed from the accreted gas, this event has occurred very long ago, and the ionized gas is 
in stable rotation on a polar orbit without forming any stars for the last many billions of years. The anomaly 
pointed out by Baes et al. \cite{baes5485} is also consistent with this fact: despite a noticeable amount of dust, 
neither neutral nor molecular hydrogen has been detected in the galaxy to a very deep limit. The large-scale stellar
disk of NGC~5485 is definitely noncircular, because the kinematical major axis at a radius greater than $25''$
is projected at a position angle $PA_0 = 100^{\circ}$, that does not coincide with the position angle of the major
axis of the isophotes. If the outer component is a disk, as suggested by the exponential surface brightness
profile, then it has an intrinsically elliptical shape and exhibits a highly noncircular stellar rotation.

\section{ACKNOWLEDGMENTS}

The long-slit spectroscopy for NGC~502 and NGC~5485 was performed in different years at the
6-m telescope under support of the SAO RAS astronomers A.V. Moiseev, V.L. Afanasiev, A.N. Burenkov,
S.N. Dodonov, and in part by courtesy of I.V. Chilingaryan; I express my sincere gratitude
to all of them. The observations at the 6-m BTA telescope are financially supported by the Ministry
of Education and Science of the Russian Federation (contract no. 14.619.21.0004, project identifier
RFMEFI61914X0004). As regards the integral-field spectroscopy, this study uses the data retrieved
from the archive of the Isaac Newton Group (of Telescopes) supported as part of the Cambridge
Astronomical Survey Unit (CASU) at the Institute of Astronomy of the Cambridge University, Great
Britain, and the data from the CALIFA (Calar Alto Legacy Integral Field Area) Survey taken from
http://califa.caha.es/. The CALIFA Survey collects data at the German–Spanish Astronomical Center
at Calar Alto (CAHA) operated jointly by the Max-Planck Society (MPG) and the Spanish National
Research Council (CSIC). This study is also based in part on public data from SDSS, SDSS-II, and
SDSS-III (http://www.sdss3.org/) financed by the Alfred P. Sloan Foundation, the institutes of the
SDSS Collaboration, the National Science Foundation, the US Department of Energy, the National
Aeronautics and Space Administration (NASA), the Japanese Monbukagakusho Foundation, the Max-Planck 
Society, and the High Education FundingCouncil for England. The research on the structure,
dynamics, and evolution of disk galaxies was supported by the Russian Science Foundation (project
no. 14-22-00041).


\clearpage

\begin{thebibliography}{}

\bibitem[1]{scorpio}
V. L. Afanasiev and A. V. Moiseev, Astron. Lett. {\bf 31}, 194 (2005).

\bibitem[2]{scorpio2}
V. L. Afanasiev and A. V. Moiseev, Baltic Astron. {\bf 20}, 363 (2011).

\bibitem[3]{dr9}
C. P. Ahn, et al., Astrophys. J. Suppl. {\bf 203}, Aid. 21 (2012).

\bibitem[4]{andersen}
D. R. Andersen, M. A. Bershady, L. S. Sparke, J. S. Gallagher III, and E. M. Wilcots, Astrophys.
J. {\bf 551}, L131 (2001).

\bibitem[5]{sauron}
R. Bacon, Y. Copin, G. Monnet, B. W. Miller, J. R. Allington-Smith, M. Bureau, C. M. Carollo,
R. L. Davies, et al., MNRAS {\bf 326}, 23 (2001).

\bibitem[6]{baes5485}
M. Baes, F. Allaert, M. Sarzi, I. de Looze, J. Fritz, G. Gentile, T.M. Hughes, I. Puerari, M.W. L. Smith,
and S. Viaene, MNRAS {\bf 444}, L90 (2014).

\bibitem[7]{atlas3d_1}
M. Cappellari, E. Emsellem, D. Krajnovic, R. M. McDermid, N. Scott, G. A. Verdoes Kleijn, L. M. Young,
K. Alatalo, et al., MNRAS {\bf 413}, 813 (2011).

\bibitem[8]{atlas3d_29}
P.-A. Duc, J.-Ch. Cuillandre, E. Karabal, M. Cappellari, K. Alatalo, L. Blitz, F. Bournaud, M. Bureau,
et al., MNRAS {\bf 446}, 120 (2015).

\bibitem[9]{franx_dz92}
M. Franx and T. de Zeeuw, Astrophys. J. {\bf 392}, L47 (1992).

\bibitem[10]{garcia}
A. M. Garcia, Astron. Astrophys. Suppl. Ser. {\bf 100}, 47 (1993).

\bibitem[11]{califa_3}
R. Garc\'ia-Benito, S. Zibetti, S. F. S\'anchez, B.Husemann, A. L. de Amorin, A. Castillo-Morales, R. Cid
Fernandes, S. C. Ellis, et al., Astron. Astrophys. {\bf 576}, Aid135 (2015).

\bibitem[12]{erwin11}
L. Guti\'errez, P. Erwin, R. Aladro, and J. E. Beckman, Astron. J. {\bf 142}, 145 (2011).

\bibitem[13]{huchra_geller}
J. P. Huchra and M. J. Geller, Astrophys. J. {\bf 257}, 423 (1982).

\bibitem[14]{n524phot}
M. A. Il'ina and O. K. Sil'chenko, Astron. Rep. {\bf 56},  578 (2012).

\bibitem[15]{atlas3d_2}
D. Krajnovic, E. Emsellem, M. Cappellari, K. Alatalo, L. Blitz, M. Bois, F. Bournaud, M. Bureau, et al.,
MNRAS {\bf 414}, 2923 (2011).

\bibitem[16]{lauri10}
E. Laurikainen, H. Salo, R. Buta, J. H. Knapen, and S. Comer\'on, MNRAS {\bf 405}, 1089 (2010).

\bibitem[17]{nirs0s}
E. Laurikainen, H. Salo, R. Buta, and J. H. Knapen, MNRAS {\bf 418}, 1452 (2011).

\bibitem[18]{mendez_abreu}
J. M\'endez-Abreu, J. A. L. Aguerri, E. M. Corsini, and E. Simonneau, Astron. Astrophys. {\bf 478}, 353 (2008).

\bibitem[19]{moiseev04}
A. V. Moiseev, J. R. Vald\'es, and V. H. Chavushyan, Astron. Astrophys. {\bf 421}, 433 (2004).

\bibitem[20]{rix_zar95}
H.-W. Rix and D. Zaritsky, Astrophys. J. {\bf 447}, 82 (1995).

\bibitem[21]{califa_1}
S. F. S\'anchez, R. C. Kennicutt, A. Gil de Paz, G. van de Ven, J. M. V\'ilchez, L. Wisotzki,
C. J. Walcher, D. Mast, et al., Astron. Astrophys. {\bf 538}, Aid. 8 (2012).

\bibitem[22]{s4g}
K. Sheth, M. Regan, J. L. Hinz, A. Gil de Paz, K. Men\'endez-Delmestre, J.-K. Mu\~noz-Mateos,
M. Seibert, T. Kim, et al., PASP {\bf 122}, 1397 (2010).

\bibitem[23]{sil05}
O. K. Sil'chenko, Astron. Lett. {\bf 31}, 227 (2005).

\bibitem[24]{sil08}
O. K. Sil'chenko, IAU Symp. 245, 277 (2008).

\bibitem[25]{silsaus0}
O. K. Sil'chenko, Astron. J. {\bf 152}, Aid. 73 (2016).

\bibitem[26]{tf_theureau}
G. Theureau, M. O. Hanski, N. Coudreau, N. Hallet, and J.-M. Martin, Astron. Astrophys. {\bf 465}, 71 (2007).

\bibitem[27]{rc3}
G. de Vaucouleurs, A. de Vaucouleurs, H. G. Corwin, Jr., R. J. Buta, G. Paturel, and P. Fouque, Third
Reference Catalogue of Bright Galaxies (Springer, Berlin, Heidelberg, New York, 1991), p. 2069.

\bibitem[28]{wagner88}
S. J. Wagner, R. Bender, and C. Moellenhoff, Astron. Astrophys. {\bf 195}, L5 (1988).


\end{thebibliography}
\end{document}